\documentclass[a4paper,11pt]{article}

\usepackage{jcappub,bm,braket,comment}
\usepackage{physics}
\usepackage[T1]{fontenc}
\usepackage{mathrsfs}
\usepackage{enumitem}
\usepackage{hyperref}
\usepackage{caption}
\usepackage[normalem]{ulem}

\newcommand{\bes}[1]{\begin{equation}\begin{split} #1 \end{split}\end{equation}}

\newcommand{\fref}[1]{fig.~\ref{#1}}
\newcommand{\Fref}[1]{Fig.~\ref{#1}}
\newcommand{\tref}[1]{tab.~\ref{#1}}

\newcommand{\sref}[1]{sec.~\ref{#1}}

\newcommand{\eref}[1]{eq.~(\ref{#1})}

\newcommand{\rref}[1]{ref.~\cite{#1}}
\newcommand{\Rref}[1]{Ref.~\cite{#1}}
\newcommand{\aref}[1]{app.~\ref{#1}}
\newcommand{\Aref}[1]{App.~\ref{#1}}

\newcommand{\Acal}{\mathcal{A}}

\newcommand{\alphaem}{\alpha_{\mathrm{em}}}
\newcommand{\cmb}{{\mathrm{cmb}}}
\newcommand{\Ndw}{N_{\mathrm{dw}}}
\newcommand{\mpl}{m_{\mathrm{pl}}}
\newcommand{\ampl}{\mathcal{A}^2\xi_0}
\newcommand{\Uniform}[2]{\mathrm{U}{({\scriptstyle #1}, {\scriptstyle #2})}}
\newcommand{\lvec}{\mathbf{l}}
\newcommand{\Lvec}{\mathbf{L}}
\newcommand{\nhat}{\hat{\mathbf{n}}}

\title{Searching for axion-like particles through CMB birefringence from string-wall networks}

\author{Mudit Jain,}
\author{Ray Hagimoto,}
\author{Andrew J. Long,}
\author{Mustafa A. Amin}
\affiliation{Department of Physics and Astronomy, Rice University, Houston, TX 77005, USA}

\emailAdd{mudit.jain@rice.edu}
\emailAdd{rmh14@rice.edu}
\emailAdd{andrewjlong@rice.edu}
\emailAdd{mustafa.a.amin@rice.edu}

\abstract{
Axion-like particles (ALPs) can form a network of cosmic strings and domain walls that survives after recombination and leads to anisotropic birefringence of the cosmic microwave background (CMB).  In addition to studying cosmic strings, we clarify and emphasize how the formation of ALP-field domain walls impacts the cosmic birefringence signal; these observations provide a unique way of probing ALPs with masses in the range $3H_0 \lesssim m_a \lesssim 3H_{\rm cmb}$.  Using measurements of CMB birefringence from several telescopes, we find no evidence for axion-defect-induced anisotropic birefringence of the CMB.  We extract constraints on the model parameters that include the ALP mass $m_a$, ALP-photon coupling $\mathcal{A} \propto g_{a\gamma\gamma} f_a$, the domain wall number $\Ndw$, and parameters characterizing the abundance and size of defects in the string-wall network.  Considering also recent evidence for isotropic CMB birefringence, we find it difficult to accommodate this with the non-detection of anisotropic birefringence under the assumption that the signal is generated by an ALP defect network.
}

\begin{document} 
\maketitle
\flushbottom

\section{Introduction}
\label{sec:intro}

Exquisite measurements of cosmic microwave background (CMB) temperature and polarization anisotropies carried out over the past few decades have revolutionized our understanding of cosmology.  The absence of $B$-mode polarization on large angular scales in the CMB has already provided important insights about cosmological initial conditions~\cite{Planck:2018jri}. Building on these measurements, more subtle analyses of achromatic CMB polarization rotation (``CMB birefringence''), has been a focus of a growing number of recent studies.  

CMB birefringence provides an exciting window into physics beyond the Standard Model. For example, hypothetical axion-like particles (ALPs) coupled to photons in the following manner
\begin{align}\label{eq:L_interaction}
    \mathscr{L}_\mathrm{int} = - \frac{1}{4} \, g_{a\gamma\gamma} \, a \, F_{\mu\nu} \tilde{F}^{\mu\nu}\;
\end{align}
can induce a birefringence signal. A photon propagating through a classical ALP field $a(x)$ is expected to experience a frequency-independent birefringence~\cite{Carroll:1989vb,Carroll:1991zs,Harari:1992ea,Carroll:1998bd} as its plane of polarization is rotated by an angle 
\begin{align}\label{eq:alpha_def}
    \alpha 
    & = - \frac{g_{a\gamma\gamma}}{2} \, \int_C \mathrm{d}X^\mu \, \partial_\mu a(X)
    \;.
\end{align}
The integral runs over the photon's worldline $X^{\mu}$ from the point of emission to detection. Isotropic birefringence has been studied in the context of an approximately homogeneous ALP field that may constitute the dark matter or dark energy~\cite{Fedderke:2019ajk,Fujita:2020aqt,Obata:2021nql,Nakatsuka:2022epj,Komatsu:2022nvu}, yielding information about the ALP-photon coupling ($g_{a\gamma\gamma}$) and the mass of the ALP ($m_a$). 

ALPs may also form a network of topological defects in the Universe, and leave a distinctive imprint on the CMB polarization via cosmological birefringence \cite{Agrawal:2019lkr,Takahashi:2020tqv,Jain:2021shf,Yin:2021kmx,Kitajima:2022jzz}. Such a defect network of strings and walls in the ALP field can exist after recombination, depending on the value of $m_a$, the number of degenerate vacua $\Ndw$, and the symmetry breaking scale $f_a$. Even if the defect network is subdominant in energy density, it can lead to a potentially detectable, anisotropic, and frequency-independent birefringence signal, which depends on $m_a$, $\Ndw$, and the anomaly coefficient $\mathcal{A}\propto g_{a\gamma\gamma}f_a$. 

Getting a handle on $m_a$, $\Ndw$, $f_a$, and $\mathcal{A}$ would be invaluable from a high energy physics point of view. For example, a global shift symmetry in the axion field would require a vanishing ALP mass $m_a=0$, whereas a nonzero mass would signal that this symmetry is broken. The general expectation is that all global symmetries are explicitly broken due to quantum gravitational effects in string theory~\cite{Kallosh:1995hi,Vafa:2005ui,Witten:2017hdv,Brennan:2017rbf,Palti:2019pca}. On the other hand, some alternative constructions of quantum gravity such as asymptotic safe gravity, may allow global symmetries to be preserved~\cite{Eichhorn:2020sbo}. Therefore, probing the mass of ALPs and the vacuum structure of their effective potential would constitute a test of the underlying nature of quantum gravity~\cite{Alvey:2021hjp}. Birefringence from the defect network probes new physics at the scale $f_a$ since the charges of particles at this scale determine the anomaly coefficient $\Acal$. Even if $f_a \gg \mathrm{TeV}$ and these particles cannot be probed directly at colliders, measurements of CMB birefringence could provide valuable insight into new high energy physics. 

\begin{figure}[t!]
	\centering
	\includegraphics
	[width=1.\textwidth]{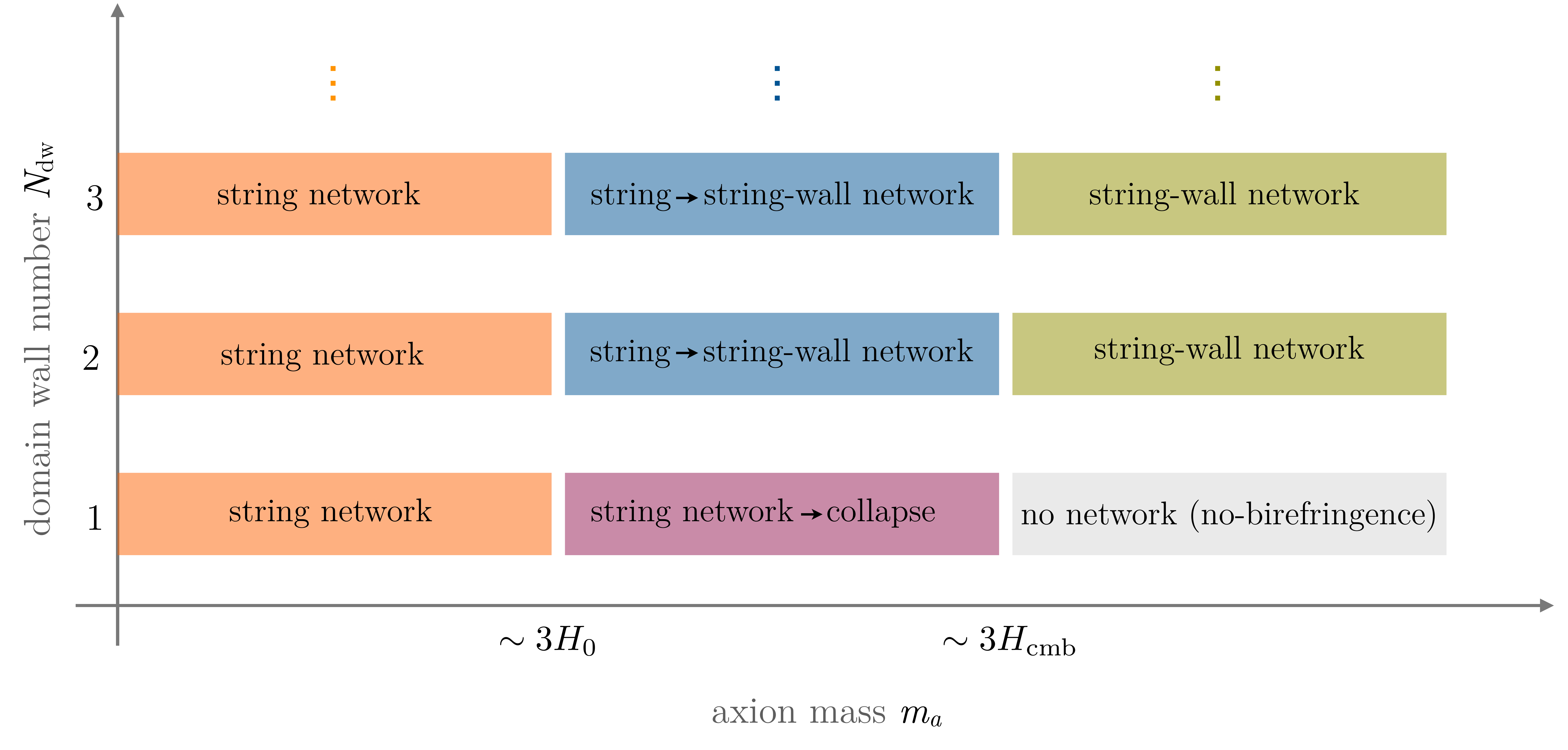} 
	\caption{\label{fig:theory_space}The types of string-wall networks seen by CMB photons travelling from the surface of last-scattering to us. The possible types of networks seen by CMB photons depend on two parameters: the domain wall number $\Ndw$ and the mass of the axion $m_a$.} 
\end{figure}

For most of this work, we focus on exceptionally light ALPs, with masses that are comparable to the Hubble parameter between recombination and today (though much higher and lower masses are also discussed). This regime is relevant for the types of string-wall networks that can be present after recombination (see \fref{fig:theory_space}). Such light masses arise naturally in many string theory constructions via non-perturbative effects~\cite{Dine:1986zy,Gibbons:1995vg,Becker:1995kb,Svrcek:2006yi,Arvanitaki:2009fg,Alvey:2020nyh}. 

The birefringence signature of an ALP defect network was first studied in \rref{Agrawal:2019lkr}. Building on that work, some of us developed semi-analytic models to calculate the expected power spectrum of axion-defect-induced birefringence for different ranges of ALP masses and network dynamics~\cite{Jain:2021shf}. Recently, \rref{Yin:2021kmx} used \textit{Planck} (2015) data to investigate some of these models (particularly those with stable string networks, orange blocks in \fref{fig:theory_space}) and derive constraints on their parameters. Taking advantage of these prior studies, our goals for the present paper are as follows:
\begin{itemize}
    \item  Provide the first constraints on collapsing ALP string-wall networks imposed by measurements of anisotropic birefringence from CMB data. 
    \item Clarify the role played by walls in string-wall networks in the context of birefringence.
    \item  Test for axion-defect-induced birefringence using measurements of anisotropic birefringence derived from data taken by various telescopes: \textit{Planck}, ACT{\sc pol}, SPT{\sc pol}, BICEP2/\textit{Keck Array}, and {\sc Polarbear}.
    \item  Validate the results of our earlier semi-analytic work on anisotropic birefringence~\cite{Jain:2021shf} by ray-tracing through statistical ensembles of defect networks and calculating the corresponding birefringence maps and power spectra.
    \item  Assess the compatibility of recent measurements of isotropic birefringence with limits on anisotropic birefringence assuming the source is an axion string-wall network.
\end{itemize}
In contrast with \rref{Yin:2021kmx}, we also investigate scenarios with $m_a\gtrsim 3H_0$ in this paper. Since walls form when $m_a \sim 3H$, this necessitates including effects due to collapsing string-wall networks or stable string-wall networks (blue, purple and green regions in \fref{fig:theory_space}). We find this higher mass window particularly intriguing since the associated phenomenology has the potential to provide a measurement of $m_a$ in the $\Ndw=1$ case. By contrast, astrophysical
observations such as probes of exotic stellar emission, are sensitive to arbitrarily light ALPs, but such observations cannot constrain their masses. 

The structure of the paper is as follows. In \sref{sec:birefringence} we review the loop-crossing model and discuss how each of the different string-wall network models are described by this framework. In \sref{sec:measurements} we briefly summarize the current status of CMB birefringence measurements. In \sref{sec:constraints} we report on the main results of our work: the non-observation of birefringence implies constraints on a hyperlight axion-like particle.  In \sref{sec:isotropic} we discuss the implications of the measurements of isotropic birefringence for our models and analysis. Finally, we summarize and conclude in \sref{sec:summary}. The article is extended by three appendices. \Aref{app:simulation} contains a detailed outline of the procedure that we have used to simulate birefringence sky maps.
\Aref{app:alpha_estimator} presents a statistical estimator that is often used to extract birefringence measurements from CMB polarization data. Finally, \aref{app:different_data} reports on searches for axion-defect-induced birefringence in additional data sets.

\begin{figure}[t!]
	\centering
	\includegraphics[width=0.9\textwidth]{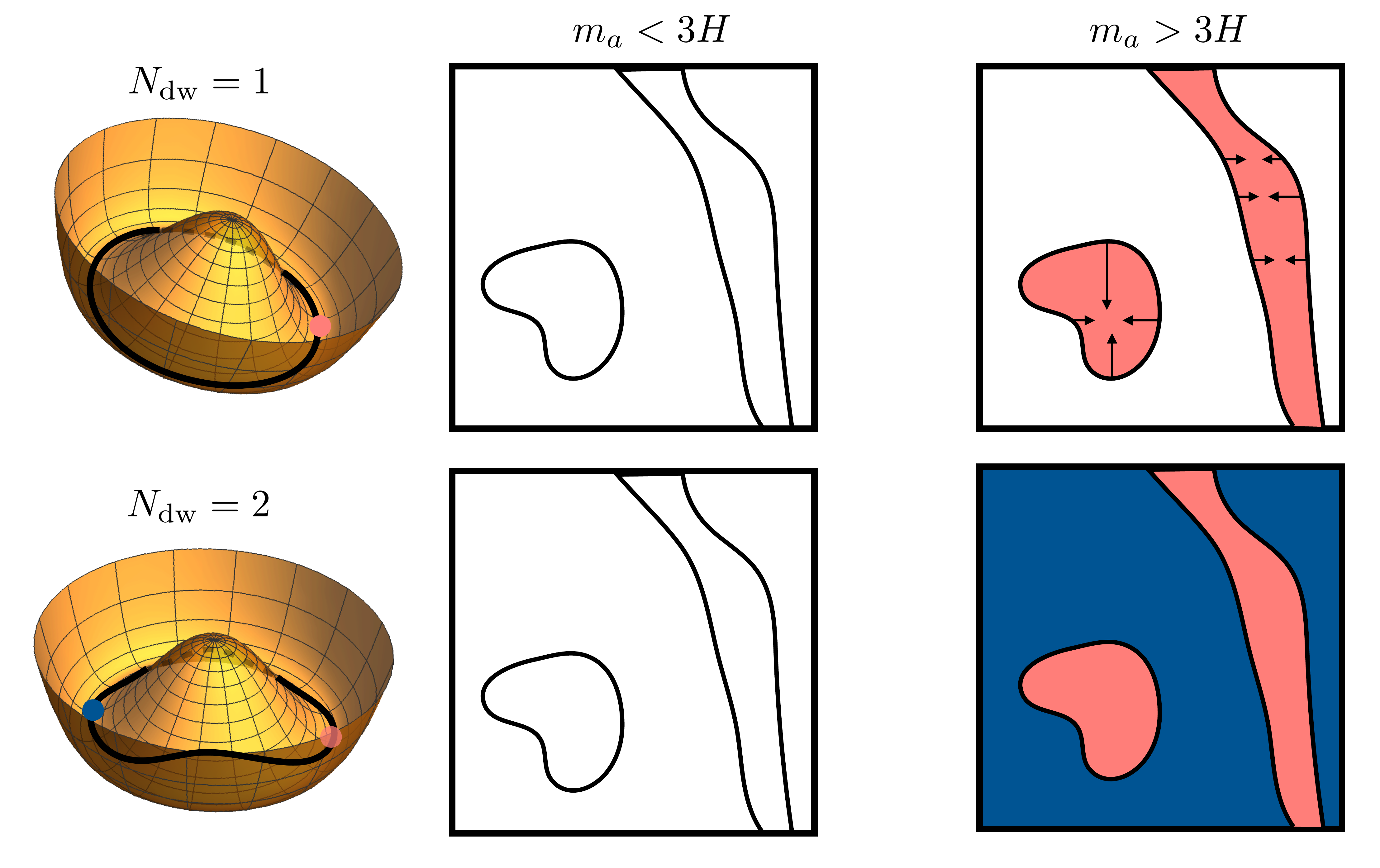}
	\caption{\label{fig:cartoon1}An illustration of the axion string-wall network dynamics -- black lines represent strings and colored regions represent walls.  {\it Top row}: For $\Ndw=1$, the string network survives from formation until $m_a \simeq 3 H(t)$. Thereafter, field gradients in the space between strings realign to form domain walls, which pull on the strings causing the network to collapse in a few Hubble times. {\it Bottom row}: For $\Ndw=2$, each string attaches to two domain walls, and the balance of forces from different walls prevents collapse and allows the network to survive after $m_a\simeq 3H$.
	}
\end{figure}

\section{CMB birefringence from an axion string-wall network}
\label{sec:birefringence}

The ALP field can form a topological defect network consisting of cosmic strings and domain walls~\cite{Vilenkin:1982ks,Vilenkin:2000jqa}.\footnote{See \rref{March-Russell:2021zfq} for an explicit discussion of cosmic strings arising in the String-Axiverse.}  
In the early Universe, if the ALP field's global symmetry is broken after inflation, then the associated phase transition fills the Universe with a network of cosmic strings~\cite{Kibble:1976sj,Kibble:1980mv}. 
The string network exhibits rich dynamics, such as the oscillation of curved string segments under the influence of their own tension, the formation of string loops from the crossing and reconnection of string segments, and the evaporation of string loops by the emission of ALPs~\cite{Sikivie:1982qv,Harari:1987ht,Chang:1998tb,Hagmann:2000ja}.  
If the Hubble parameter drops below the axion mass scale, $3H(t) \simeq m_a$, the ALP field in the space between strings is released from Hubble drag, and the strings become bounded by domain walls; see \fref{fig:cartoon1} for an illustration. 
The number $\Ndw$ of domain walls attached to each string is a parameter of the theory, associated with explicit symmetry breaking.  
For models with $\Ndw = 1$ the string-wall network collapses into a bath of ALPs within a few Hubble times, but for $\Ndw \geq 2$ the network is stable due to the balance of forces. 

To assess the implications of an ALP string-wall network for CMB birefringence, it is useful to break up the parameter subspace ($m_a, \Ndw$) into four regions as illustrated in \fref{fig:theory_space}.  
For $m_a \lesssim 3 H_0$, the domain walls have not yet formed by today, which makes $\Ndw$ irrelevant, and the birefringence signal arises from axion strings alone.  
For $3 H_0 \lesssim m_a \lesssim 3 H_{\cmb}$ and $\Ndw = 1$, the domain walls form and cause the network to collapse between recombination and today. 
This shuts off the accumulation of birefringence at the time when $3H(t) \simeq m_a$.  
For $3 H_0 \lesssim m_a \lesssim 3 H_{\cmb}$ and $\Ndw \geq 2$, the formation of domain walls converts the string network into a stable string-wall network between recombination and today, whereas for $3 H_{\cmb} \lesssim m_a$ and $\Ndw \geq 2$, this conversion occurs before recombination. 

\begin{figure}[t!]
	\centering
	\includegraphics[width=0.95\textwidth]{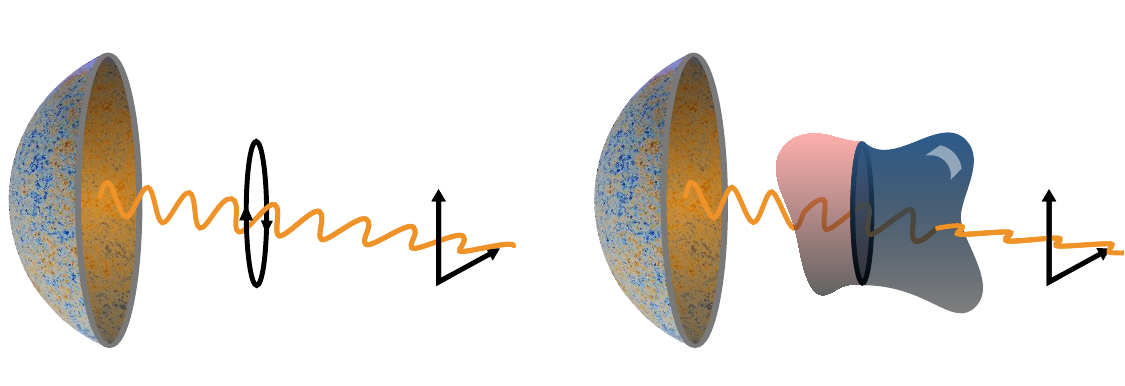} 
	\caption{\label{fig:cartoon2}An illustration of axion-defect-induced birefringence.  \textit{Left:}  A photon's plane of polarization rotates gradually as it approaches and passes through a string loop without domain walls.  \textit{Right:}  For a string loop bounded by two domain walls, the polarization axis rotates `abruptly' upon crossing each wall.  In both scenarios, with and without domain walls, the net effect is the same rotation angle $\alpha = \pm \Acal \alphaem$. }
\end{figure}

When coupled to electromagnetism \eqref{eq:L_interaction}, ALP strings and walls induce a frequency-independent birefringence signal. 
This signal is insensitive to the symmetry breaking scale $f_a$ but directly probes the anomaly coefficient $\Acal = - \pi f_a g_{a\gamma\gamma} / \alphaem$~\cite{Agrawal:2019lkr}. 
For instance, if walls have not yet formed, then a photon crossing through a string loop `sees' the axion field pass through a full cycle $\Delta a = \int_C \mathrm{d}X^\mu \, \partial_\mu a(X) \to \pm 2\pi f_a$, and the resultant birefringence angle \eqref{eq:alpha_def} is $\alpha \to \pm \Acal\alphaem$. 
The $\pm$ factor is the string's winding number,\footnote{Although strings can have winding numbers $\pm 1, \pm 2, ...$, only the $\pm 1$ strings are the most stable. Throughout our work, we only consider $+1$ or $-1$ winding numbers.} and the arrow indicates the limiting value as the end points of $C$ are taken far away from the loop; see \fref{fig:cartoon2} for an illustration. 
If domain walls are present in the network, the birefringence effect is $\alpha = \pm \Acal\alphaem/\Ndw$ at each wall crossing, since the axion field changes ``abruptly'' by $\Delta a = \pm 2\pi f_a/\Ndw$.  
However, since each string connects to $\Ndw$ walls, the net effect is insensitive to $\Ndw$; we discuss this point further in \sref{sub:stable_string_wall}.  

To calculate the birefringence signal from an axion string-wall network, we employ the `loop crossing model' developed in refs.~\cite{Agrawal:2019lkr,Jain:2021shf}.  
The loop crossing model captures features of the network's rich structure and dynamics that are particularly relevant for birefringence.  
In this framework the network is treated as a collection of circular, planar string loops that are uniformly distributed throughout space and isotropically oriented. 
The density of loops and its time dependence are controlled by the model parameters.  
The birefringence is calculated by considering photons propagating through the network and associating an angle $\alpha = \pm \Acal\alphaem$ to a photon that traverses the disk bounded by a loop.  
As the photon crosses through multiple loops with random winding numbers $\pm 1$, the accumulated phase-shift grows like a random walk.  
For two points on the sky $\hat{\gamma}_1$ and $\hat{\gamma}_2$ separated by an opening angle $\theta_o = \arccos(\hat{\gamma}_1\cdot\hat{\gamma}_2)$, the correlation between the accumulated birefringence of CMB photons from these points is taken to be~\cite{Agrawal:2019lkr} 
\begin{align}\label{eq:DPhi_from_Nboth}
    \langle \alpha(\hat{\gamma}_1) \, \alpha(\hat{\gamma}_2) \rangle = (\Acal\alphaem)^2 \, N_\text{both}(\theta_o) 
    \;,
\end{align}
where $N_{\mathrm{both}}(\theta_o)$ is the average number of loops that both photons traverse.
The associated angular power spectrum $C_\ell^{\alpha\alpha}$ is given by 
\begin{equation}\label{eq:Cl_def}
    C_\ell^{\alpha\alpha} = 2\pi \int^{1}_{-1} \mathrm{d}(\cos\theta_o) \, P_\ell(\cos\theta_o) \, \langle\alpha(\hat{\gamma}_1)\alpha(\hat{\gamma}_2)\rangle\;,
\end{equation}
and $\ell(\ell+1)C^{\alpha\alpha}_{\ell}/2\pi$ would be constant for scale-invariant anisotropic birefringence.  

\begin{figure}[t]
	\centering
    \includegraphics[width=6in]{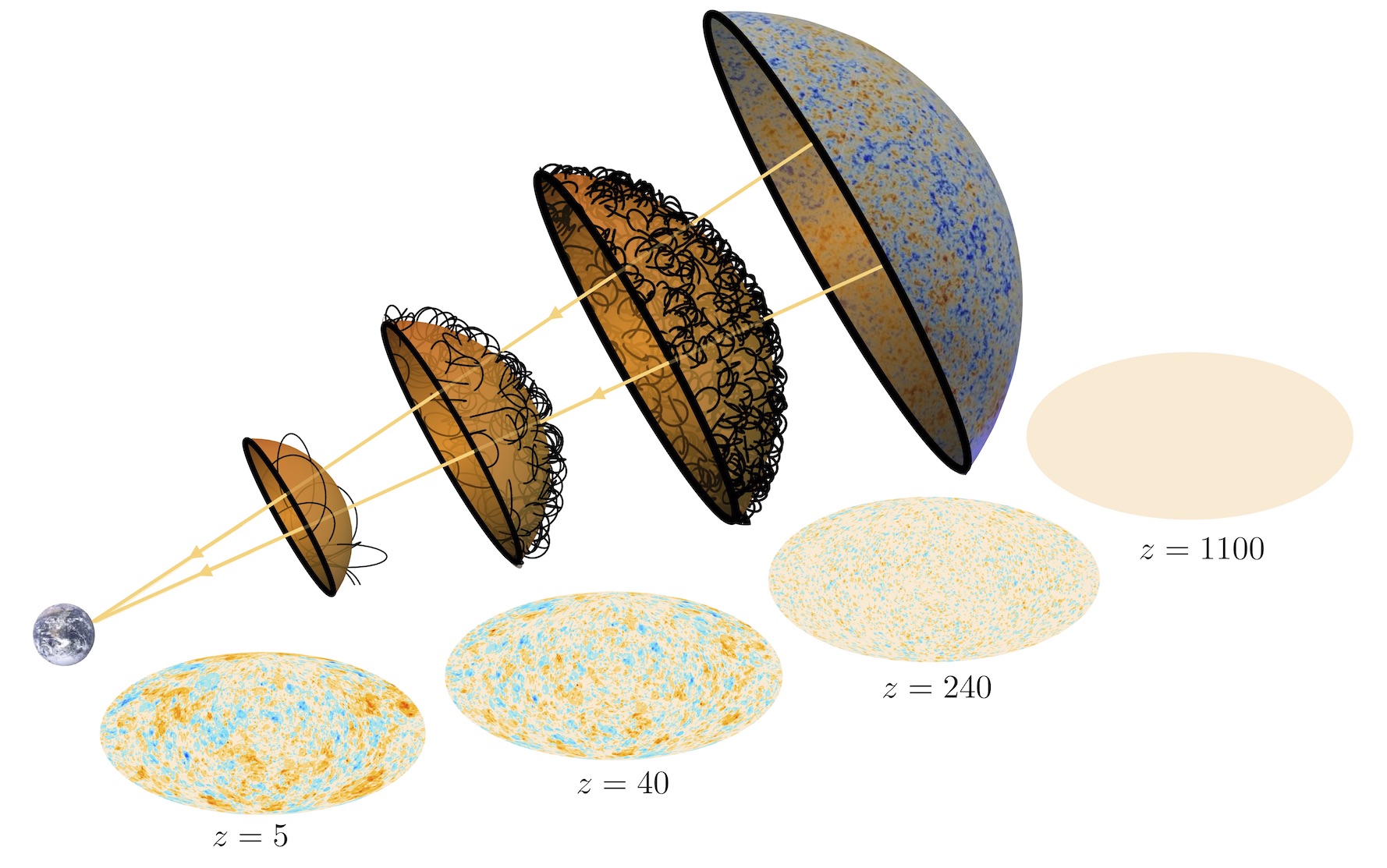}
	\caption{\label{fig:LCM}An illustration of the loop-crossing model that we use to calculate the axion-defect-induced birefringence signal.  The string-wall network is modeled as a collection of randomly oriented circular loops. The abundance and radius of the loops evolves in time, tracking the Hubble scale; parameter $\xi_0$ controls the number of loops per Hubble volume and parameter $\zeta_0$ controls the loop size in Hubble units. Photons crossing through a loop experience birefringence $\alpha = \pm \mathcal{A} \, \alphaem$, and multiple loop crossings add incoherently like a random walk. The all-sky maps show the birefringence angle $\alpha$ that has accumulated up to the stated redshift based on a typical realization of the defect network.}
\end{figure}

The model dependence enters through $N_{\mathrm{both}}$, which knows about the density of loops in the network and their length distribution.  
It can be analytically approximated as~\cite{Jain:2021shf}
\begin{align}
    N_{\mathrm{both}}(\theta_o) \approx \int^{\infty}_{0}\!\mathrm{d}\zeta\int^{\tilde{z}_{\ast}(\zeta,\theta_o)}_{0}\!\mathrm{d}z\;Q(\zeta,z,0)\,\chi(\zeta,z)\;,
\end{align}
where $\zeta$ is a dimensionless measure of loop length, $z$ is redshift, $Q(\zeta,z,0)$ is a kernel function, and $\chi(\zeta,z)$ contains the model dependence.  
Expressions for $Q(\zeta,z,0)$ and $\tilde{z}_{\ast}(\zeta,\theta_o)$ are available in eqs.~(3.25)~and~(3.34) of \rref{Jain:2021shf}.  
For the networks that we study in the following subsections, the defects' size tracks the growing Hubble scale.  
Consequently birefringence on small angular scales is imprinted by small loops at early times, whereas large-scale features are imprinted at later times; see \fref{fig:LCM} for an illustration.  
The model function $\chi(\zeta,z)$ is normalized such that it becomes time-independent for a string network in scaling $\chi(\zeta,z) = \chi(\zeta)$, and then the average string length per Hubble volume is $\xi_0/H$ with constant $\xi_0 = \int_0^\infty \! \mathrm{d} \zeta \ \chi(\zeta)$.  

Recent numerical simulations have sparked some debate as to whether global string networks (such as the ones we consider here) exhibit scaling~\cite{Yamaguchi:1998gx,Yamaguchi:2002sh,Hiramatsu:2010yu,Hiramatsu:2012gg,Kawasaki:2014sqa,Lopez-Eiguren:2017dmc,Hindmarsh:2019csc,Hindmarsh:2021vih} with constant $\xi_0$ or whether they deviate from scaling~\cite{Gorghetto:2018myk,Vaquero:2018tib,Kawasaki:2018bzv,Martins:2018dqg,Buschmann:2019icd,Klaer:2019fxc,Gorghetto:2020qws} with a slowly-growing $\xi_0$.  
In our work, the CMB birefringence signal only depends upon the string network evolution between recombination and today, so a logarithmic change in $\xi_0$ would induce a $1 - (\log f_a/H_\cmb)/(\log f_a/H_0) \sim \mathcal{O}(10\%)$ effect on the birefringence signal, which can be neglected.  
If the network maintains scaling we expect $\xi_0 \sim \mathcal{O}(\mathrm{few})$ and if there is a logarithmic growth during the long time interval between formation and recombination, we expect $\xi_0 \sim \log f_a/H_\cmb \sim \mathcal{O}(10)$; our analysis captures both scenarios by setting a wide prior on $\xi_0$.  

\subsection{Stable string network}
\label{sub:stable_strings}

For sufficiently small ALP masses $m_a < 3H_0$ and arbitrary $\Ndw$, domain walls have not yet formed today, and the defect network is a stable string network.  
We consider the `uniform loop size' model of \rref{Jain:2021shf}; the model function is 
\begin{align}\label{eq:chi_model_1}
    \chi(\zeta) = \xi_0 \, \delta(\zeta - \zeta_0) 
    \;,
\end{align}
such that all loops in the network at time $t$ have the same radius $\zeta_0 / H(t)$, and $\xi_0$ is the average number of loops per Hubble volume.  
Recent simulations~\cite{Gorghetto:2018myk,Vaquero:2018tib,Kawasaki:2018bzv,Martins:2018dqg,Buschmann:2019icd,Klaer:2019fxc,Gorghetto:2020qws} of axion string networks identify a dominant population of large loops and infinite strings, which motivates parameters: $\xi_0 = \mathcal{O}(1-10)$ and $\zeta_0 = \mathcal{O}(0.1-1)$.  
The birefringence two-point correlation function is approximately~\cite{Jain:2021shf}
\begin{align}\label{eq:DPhi_model_1}
    \langle \alpha(\hat{\gamma}_1) \, \alpha(\hat{\gamma}_2) \rangle
    \approx \ampl \, \alphaem^2 
    \begin{cases}
    \dfrac{\zeta_0}{4} \, 
    \left( \log\bigl(1+\tilde{z}_\ast(\zeta_0,\theta_o)\bigr) - \dfrac{\zeta_0}{3} \right) & \qquad \theta_o < \theta_t\\
    \dfrac{1}{3\zeta_0} \, 
    \log^3\bigl(1+\tilde{z}_\ast(\zeta_0,\theta_o)\bigr) &
    \qquad \theta_o > \theta_t 
    \end{cases}
    \;,
\end{align}
where $\theta_t \approx 1$. 
Note that there is a degeneracy between $\xi_0$ and $\mathcal{A}$; only the combination $\ampl$ appears in the correlation function, and it controls the amplitude of the signal.  

\begin{figure}[t]
    \centering
    \raisebox{-0.5\height}{\includegraphics{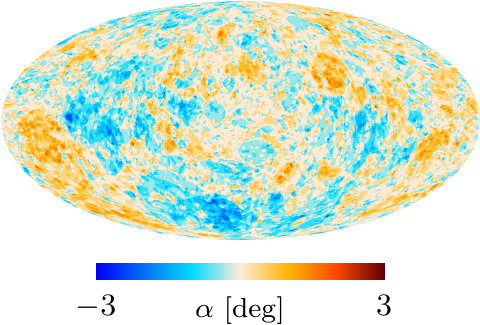}}\hspace{0.05in}
    \raisebox{-0.5\height}{\includegraphics{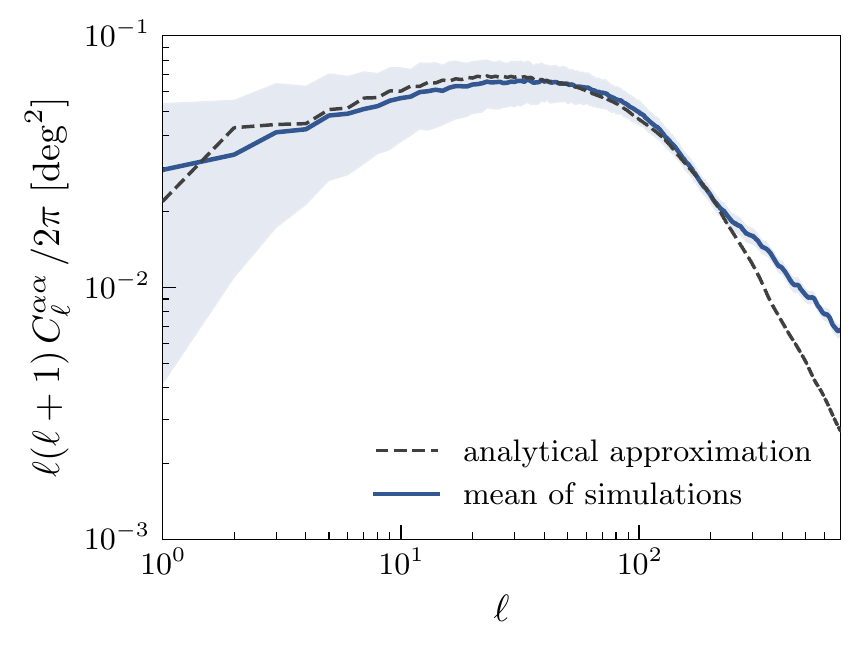}}
	\caption{\label{fig:signal_strings}The expected birefringence signal due to a string network that survives until today. We take $\ampl = 1$ and $\zeta_0 = 1$.  \textit{Left:}  A sample sky map of the birefringence angle $\alpha(\nhat)$. \textit{Right:}  The angular power spectrum of the birefringence angle $C_\ell^{\alpha\alpha}$.  The dashed black curve is our analytical approximation~\eqref{eq:DPhi_model_1}, while the blue curve is the mean of a suite of 1000 simulations of the loop crossing model. The shaded region shows the corresponding $68\%$ central containment interval exhibiting cosmic variance.}
\end{figure}

The expected birefringence signal is shown in \fref{fig:signal_strings} for $\zeta_0 = 1$ and $\ampl = 1$.  
The left panel shows a simulated realization of the birefringence angle over the sky; see \aref{app:simulation} for details of the simulation.  
Note the loop-like features that span a wide range of angular scales.  
The right panel shows the corresponding angular power spectrum \eqref{eq:Cl_def}.  
For each realization we calculate a $C_\ell^{\alpha\alpha}$; the blue curve shows their mean and the blue band shows the $68\%$ containment region.  
This band broadens toward low $\ell$ due to cosmic variance.  
Note that the spectrum is almost scale invariant for $\ell \lesssim 100$, which follows from the assumed scale invariance of the string network.  
Exact scale invariance is broken by the angular size of the string loops at recombination, which sets a minimal angular scale for the birefringence anisotropies that corresponds to the peak at $\ell_p \sim 0.1\pi/(\zeta_0\,\theta_\cmb) \approx 40/\zeta_0$.  
The right panel also shows the analytic approximation in \eref{eq:DPhi_model_1} as the gray-dashed curve.  
Note that the approximation agrees exceptionally well with the direct simulation, which partly validates our use of \eref{eq:DPhi_model_1} for data fitting and parameter constraints in \sref{sec:constraints}.  

\subsection{Collapsing string-wall network}
\label{sub:collapsing_string_wall}

For ALP mass in the range $3H_0 < m_a < 3H_\cmb$ with $\Ndw = 1$, the string network develops domain walls and collapses between recombination and today.  
We consider the `string network collapse' model of~\cite{Jain:2021shf}, which has the model function 
\begin{align}\label{eq:chi_model_3}
    \chi(\zeta,z) = \xi_0 \ \delta(\zeta - \zeta_0) \ \Theta(z - z_c) 
    \;.  
\end{align}
The dimensionless parameters $\xi_0$ and $\zeta_0$ have the same interpretation as in \sref{sub:stable_strings}, and $z_c$ is the redshift when $3H(t) = m_a$, which is given by 
\begin{align}\label{eq:zc_from_ma}
    z_c = \left[\left(\frac{(m_a/3H_0)^2 - \Omega_{\Lambda}}{\Omega_{m}}\right)^{1/3}-1 \right]\;,
\end{align}
in $\Lambda\mathrm{CDM}$ cosmology. 
The step function $\Theta(z-z_c)$ models a rapid formation of domain walls at $z=z_c$ and an abrupt collapse of the string-wall network, which shuts off any further accumulation of birefringence.\footnote{After the string-wall network collapses, its energy is transferred to a population of non-relativistic ALPs (a subdominant component of the dark matter), which continue to induce birefringence.  However, this contribution to the total birefringence is suppressed at low $\ell$ by $\sim 10^{-3} H_0 / m_a$ for $m_a \gtrsim 100 H_0$, making it negligible~\cite{Jain:2021shf}.}
The birefringence two-point correlation function is approximately given as~\cite{Jain:2021shf}
\begin{align}\label{eq:DPhi_model_3}
    \langle \alpha(\hat{\gamma}_1) \, \alpha(\hat{\gamma}_2) \rangle
    \approx \xi_0\bigl( \Acal \, \alphaem \bigr)^2
    \begin{cases}
    \frac{\zeta_0}{4}
    \log\left(\frac{1+(z_c^{3/2} + \tilde{z}^{3/2}_\ast)^{2/3}}{1+z_c}\right) \qquad &\theta_o, \theta_c < \theta_t\\
    \frac{\zeta_0}{4}
    \log\left(1+(z_c^{3/2} + \tilde{z}^{3/2}_\ast)^{2/3}\right)\\
    \qquad - \frac{1}{3\zeta_0}\log^3(1+z_c) - \frac{\zeta_0^2}{12} \qquad &\theta_o < \theta_t < \theta_c\\
    \frac{1}{3\zeta_0}\Bigl(\log^3(1+(z_c^{3/2} + \tilde{z}^{3/2}_\ast)^{2/3})\\
    \qquad\qquad - \log^3(1+z_c)\Bigr) \qquad & \theta_t < \theta_o, \theta_c 
    \end{cases}
    \;,
\end{align}
where $\theta_c$ corresponds to the effective angular size of loops at the time of collapse. 

\begin{figure}[t]
	\centering
    \raisebox{-0.425\height}{\includegraphics{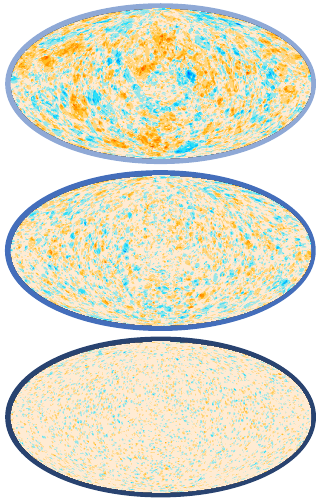}}\hspace{0.05in}
    \raisebox{-0.5\height}{\includegraphics{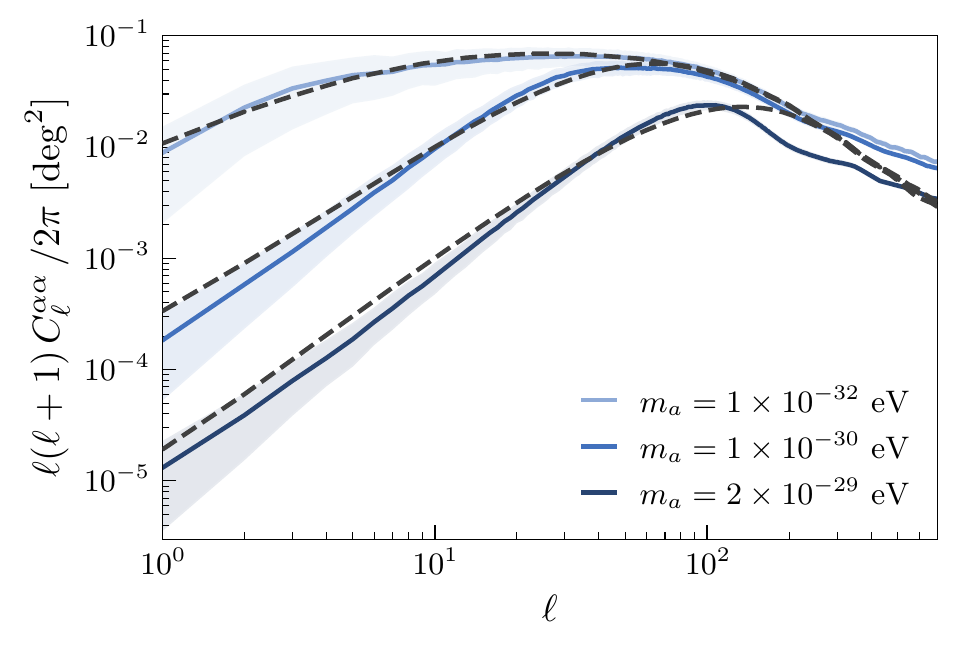}}
	\caption{\label{fig:signal_collapse}Same as \fref{fig:signal_strings} but for a collapsing string-wall network ($\Ndw = 1$).  Several curves for different axion mass $m_a$ are shown, corresponding to different collapse redshifts $z_c$ given by \eref{eq:zc_from_ma}. For each $m_a$, we also show the corresponding all-sky birefringence map on the left. Increasing $m_a$ causes the network to collapse earlier, and suppresses power at small $\ell$ (large angular scales).}
\end{figure}

We show the expected birefringence signal in \fref{fig:signal_collapse} for $\ampl = 1$, $\zeta_0 = 1$, and three choices of the ALP mass $m_a$.  
Note that $3H_0 \approx 4.5 \times 10^{-33} \ \mathrm{eV}$ (for $h=0.7$) and $H_\mathrm{cmb} \approx 1.0 \times 10^{-29} \ \mathrm{eV}$ (for $z_\mathrm{cmb} = 1100$, $\Omega_m = 0.3$, $\Omega_\Lambda = 0.7$, $\Omega_r = 9 \times 10^{-5}$).  
Raising the ALP mass causes the network to collapse earlier, which suppresses power at large angular scales, since larger loops would have formed later.  
The power spectrum displays a strong scale dependence $\ell (\ell+1) C_\ell^{\alpha\alpha} \propto \ell^2$ for $\ell\lesssim \ell_c \sim \pi/\theta_c(m_a)$, where $\theta_c(m_a)$ corresponds to the angular size of loops at the time of network collapse~\cite{Jain:2021shf}.  
The power spectrum calculations also show good agreement between the direct numerical simulation and the analytic approximation in \eref{eq:DPhi_model_3}.
An $\mathcal{O}(1)$ discrepancy develops at large $m_a$ for high $\ell \gtrsim 100$; for the purpose of data analysis, we neglect this mismatch and use the analytic calculation (dashed curves).  

\subsection{Stable string-wall network}
\label{sub:stable_string_wall}

For $\Ndw \geq 2$ the formation of domain walls at the time when $3H(t) \simeq m_a$ leads to a stable string-wall defect network.  
The resultant CMB birefringence signal is expected to be qualitatively unchanged from the stable string network without walls~\cite{Agrawal:2019lkr}, which we discussed already in \sref{sub:stable_strings}.  
We argue this point in two steps: first, we argue that the realignment of smooth axion field gradients around strings into sharp gradients across walls does not change the integrated gradient `seen' by a photon propagating through the network; and second, we argue that the abundance of walls in the network follows the same scaling as the abundance of string loops in the wall-free network.  

\begin{figure}[t!]
	\centering
	\includegraphics[width=6.1in]{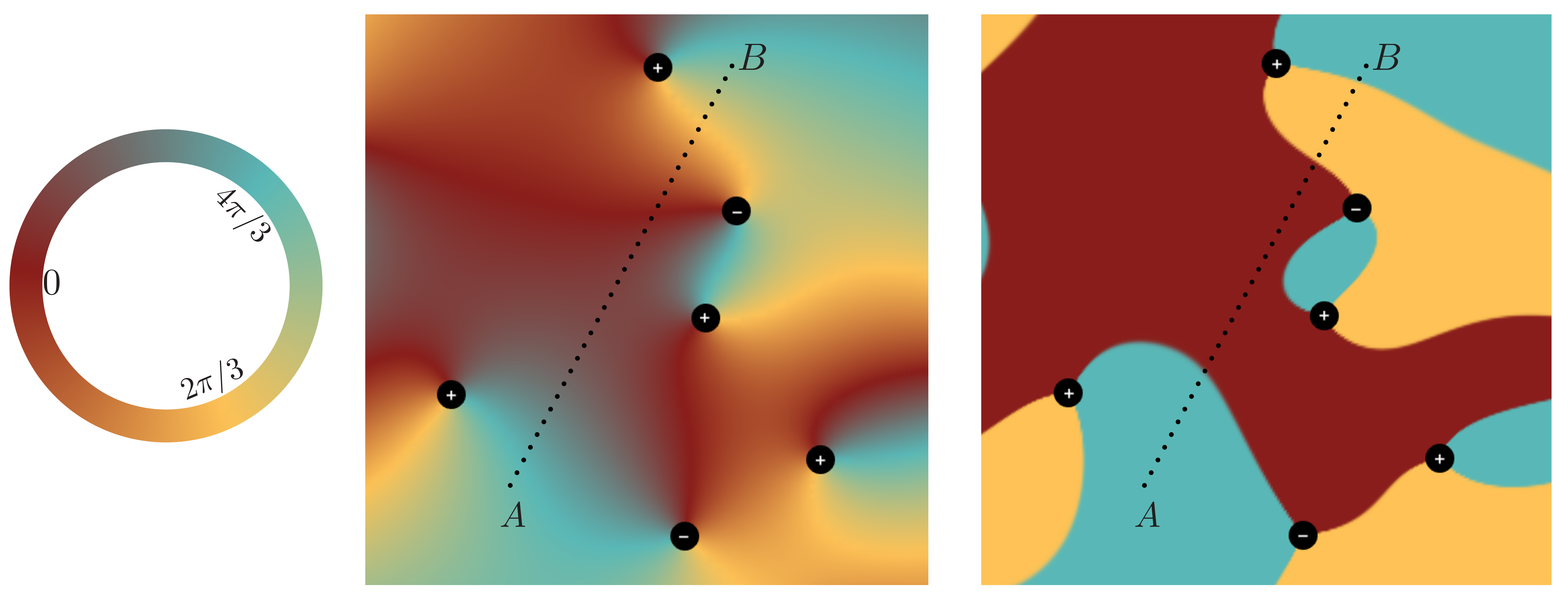}
	\caption{\label{fig:stable.string.wall.cartoon}An illustration of the axion field $a/f_a \in [0,2\pi)$ in the vicinity of several parallel long strings (vortices) with and without walls. \textit{Left panel:} Color legend indicating axion field values on a circle. \textit{Middle panel:} The field around a long string in cylindrical coordinates obeys ${\bm \nabla} a = \pm \hat{\bm \theta}/\rho$, and several strings are superimposed to form the middle image. \textit{Right panel:} Long strings are connected to $\Ndw = 3$ domain walls corresponding to sharp field gradients between minima at $a/f_a = 0, 2\pi/3$, and $4\pi/3$. }
\end{figure}

Since the birefringence signal is proportional to the integrated axion field gradient~\eqref{eq:alpha_def}, it is necessary to understand how this quantity differs whether or not walls are present in the network.  
For a network of axion strings without domain walls, the axion field's gradient varies smoothly throughout space.  
We illustrate this behavior in \fref{fig:stable.string.wall.cartoon} by showing an axion field configuration for a collection of parallel long strings (vortices in two dimensions).  
For instance, along the path from point $A$ to point $B$ the axion field passes through a full cycle and the integrated field gradient is $\Delta a = 2 \pi f_a$.  
If the same collection of strings were each connected to $\Ndw = 3$ domain walls, then the field gradients would be localized in space in order to minimize the energy of the configuration.  
Nevertheless, on the path from $A$ to $B$ the integrated field gradient would remain equal to $\Delta a = 2 \pi f_a$; each wall contributes only $2 \pi f_a / 3$, but there are $3$ walls along the path.  
More generally, the presence or absence of walls for a given collection of vortices will not impact the integrated phase gradient modulo sub-$2\pi$ variation in the field value at the endpoints.  

The birefringence signal also depends upon the abundance of domain walls in the string-wall network.  
Here we argue that for a network in scaling, the number density of domain walls tracks the number density of string loops, which evolves in the same way whether or not there are walls.  
Since every domain wall ends on a string and each string has $\Ndw$ domain walls then the number densities of walls and strings are related by $n_\mathrm{dw} \approx \Ndw n_s$.\footnote{Walls may also close on themselves forming bubbles.  A photon passing through one of these configurations does not experience a net birefringence since the contributions from the two wall crossings cancel.  On the other hand, some component of the birefringence signal arises from photons emitted within a bubble and detected at a point outside (or vice versa)~\cite{Kitajima:2022jzz}.  Assuming that bubbles are not nested, this component of the birefringence signal may be as large as $|\alpha| = \Acal \alphaem / \Ndw$.  However, for the string-wall networks that we consider, the effect of multiple wall crossings (loop crossing) allows $|\alpha|$ to accumulate to values that are larger than $\Acal \alphaem$, which is the dominant component. }
Numerical simulations~\cite{Ryden:1989vj,Hiramatsu:2010yn,Hiramatsu:2012sc} of axion string-wall networks indicate that the strings reach a scaling regime in which the number density $n_s(t)$ evolves as if the walls were absent.  
Thus we conclude that the density of walls in the string-wall network tracks the density of strings in the wall-free network.  

It is worth noting that these arguments apply equally well for models with $m_a > 3 H_\cmb$ that form walls before recombination and for those with $3 H_0 < m_a < 3H_\cmb$ that form walls after recombination.  
In terms of the parameter space shown in \fref{fig:theory_space}, the CMB birefringence signal will be qualitatively unchanged for ``string network,'' ``string$\to$string-wall network,'' and ``string-wall network.''  

The energy density in the stable string-wall network is stored mostly in the rest mass of the domain walls, which have tension $\sigma \approx 8 m_a f_a^2 / \Ndw^2$.  
Since the wall's energy redshifts more slowly than strings, the stable string-wall network could present a problem for cosmological observables.  
If $\xi_\mathrm{dw}$ is the average number of walls per Hubble volume today, then the string-wall network's energy density is approximately $\xi_\mathrm{dw} \sigma H_0$.  
A weak cosmological constraint is obtained by requiring this energy to be small compared to the critical density today $3 \mpl^2 H_0^2$, which implies:  
\begin{align}\label{eq:overclosure}
    \frac{\xi_\mathrm{dw} \sigma H_0}{3\mpl^2 H_0^2} 
    \sim 
    \biggl( \frac{\xi_\mathrm{dw}}{\Ndw/2} \biggr) 
    \biggl(\frac{\Ndw}{2}\biggr)^{-1}
    \biggl(\frac{m_a}{10^{-20} \,\mathrm{eV}}\biggr) 
    \biggl(\frac{f_a}{10^{12}\ \mathrm{GeV}}\biggr)^2 
    \ll 1 
    \;.
\end{align}
It is worth emphasizing that a sufficiently small decay constant $f_a$ allows a larger axion mass $m_a$ while still satisfying the overclosure condition and observational constraints on the axion-photon coupling (non-minimal models such as clockwork axion~\cite{Choi:2015fiu,Kaplan:2015fuy} may help to further open this parameter space).  
For instance, if $f_a \sim 10^{10} \ \mathrm{GeV}$ (allowed in type IIB String Theory constructions in the large volume scenario~\cite{Balasubramanian:2005zx,Conlon:2005ki}), then the upper bound on $m_a$ can even be as large as $\sim 10^{-16} \ \mathrm{eV}$ with $\xi_\mathrm{dw}/\Ndw \sim \mathcal{O}(1)$. 
Axion masses above $m_a \gtrsim 10^{-21} \ \mathrm{eV}$ are especially interesting, since they avoid the Lyman-$\alpha$ bound on ultralight dark matter~\cite{Hui:1998hq,Irsic:2017yje}.  
This scenario in which the ultralight ALP makes up a significant fraction of dark matter and leads to a birefringence signal in the CMB merits closer study, which we do not pursue further here.  

\section{Measurements of cosmological birefringence with CMB data}
\label{sec:measurements}

Cosmological birefringence is expected to leave a distinctive imprint on the polarization anisotropies in the cosmic microwave background radiation.  
At a given point on the sky, a rotation in the CMB's plane of polarization cannot be measured directly, since the initial orientation on the surface of last scattering is not known.  
Nevertheless, the polarization pattern across the sky carries information about cosmological birefringence.  
CMB polarization maps may be decomposed into parity-even $E$-mode and parity-odd $B$-mode type polarization patterns~\cite{Kamionkowski:1996zd,Seljak:1996gy,Kamionkowski:1996ks}.  
Thompson scattering at the surface of last scattering generates $E$-mode polarization, whereas $B$-mode polarization requires parity-violating sources, such as gravitational wave radiation~\cite{Abbott:1984fp}.  
Cosmological birefringence partially converts $E$-mode polarization into $B$-mode polarization (and vice versa).  
This induces a $B$-mode signal if none was present otherwise and leads to correlations among the temperature and polarization patterns~\cite{Kamionkowski:2008fp,Yadav:2009eb,Gluscevic:2009mm}. 

In order to extract information about birefringence from CMB temperature and polarization data, it is customary to work with a set of statistical quantities that provide unbiased estimators of the birefringence~\cite{Kamionkowski:2008fp,Yadav:2009eb,Gluscevic:2009mm}.  
These $\alpha$-estimators are constructed from pairs of CMB power spectra (possibly also correlating across different frequency bins).  
The power spectrum of the $\alpha$-estimators are equal to the birefringence angular power spectrum $C_\ell^{\alpha\alpha}$ up to a noise term.  
In this way, the measured $EE$, $BB$, $EB$, $TE$, and $TB$ power spectra are used to reconstruct the birefringence power spectrum. 
For pedagogical purposes, in \aref{app:alpha_estimator} we define the $\alpha$-estimators and demonstrate the reconstruction procedure for mock polarization data.

\begin{figure}[t]
	\centering
	\includegraphics{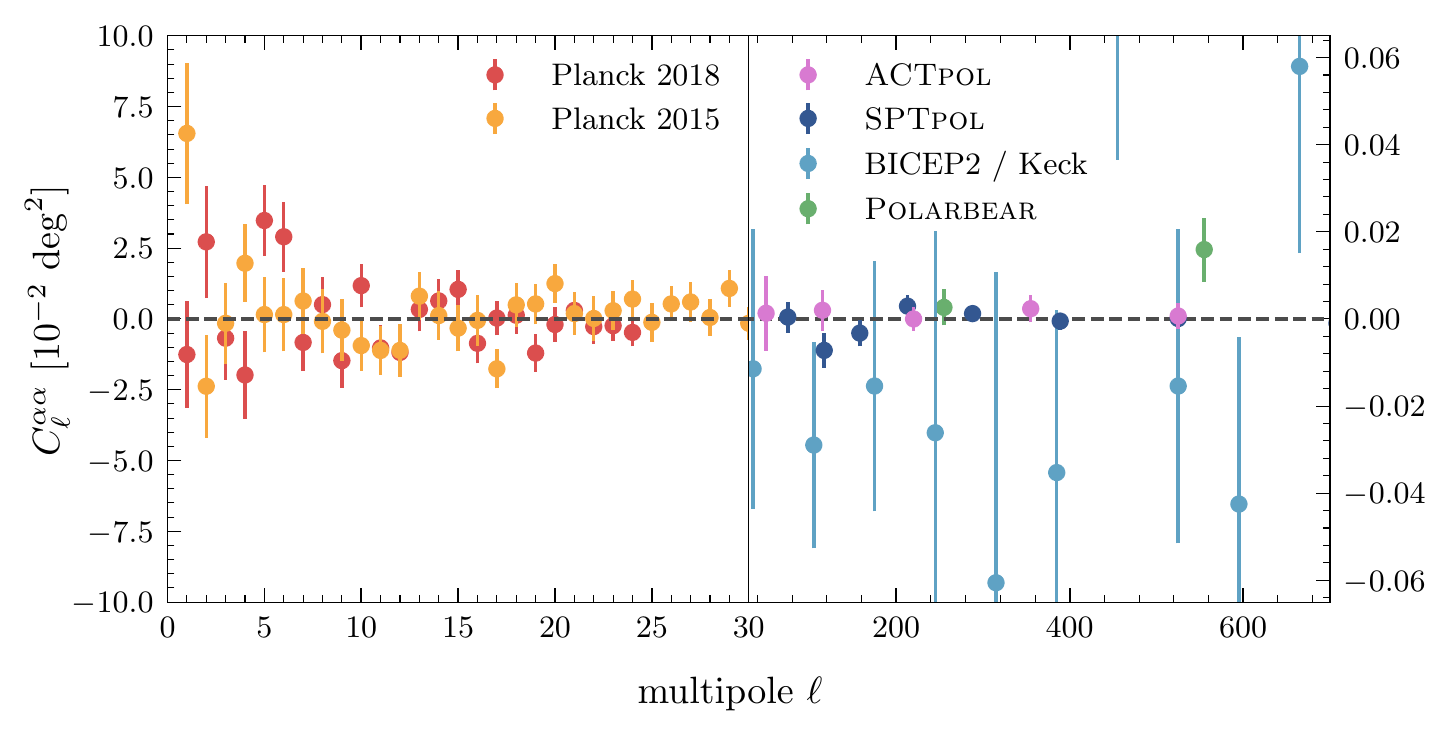}
	\caption{\label{fig:data}Measurements of anisotropic cosmological birefringence with data from various CMB telescopes: \textit{Planck} (2018)~\cite{Bortolami:2022whx} (see also \rref{Gruppuso:2020kfy}), \textit{Planck} (2015)~\cite{Contreras:2017sgi}, ACT{\sc pol}~\cite{Namikawa:2020ffr}, SPT{\sc pol}~\cite{SPT:2020cxx}, BICEP2/\textit{Keck Array} \cite{BICEP2:2017lpa}, and {\sc Polarbear}~\cite{POLARBEAR:2015ktq}. Note the different scales for $\ell < 30$ and $\ell > 30$.  We do not show the measurements from {\it Planck} (2015) for $\ell > 30$ due to the large error bars.  Additional data is available at higher multipoles, which is also not shown in this summary figure.  These measurements are consistent with the absence of anisotropic cosmological birefringence.  
	}
\end{figure}

The data from various CMB telescopes have been analyzed to search for evidence of cosmological birefringence.  
Assuming that the birefringence effect is isotropic (same rotation angle $\alpha$ at every point on the sky), several studies have recently reported evidence for nonzero birefringence~\cite{Minami:2020odp,Diego-Palazuelos:2022dsq,Eskilt:2022wav,Eskilt:2022cff}, including \rref{Eskilt:2022cff} that reports an angle ${0.342^\circ}_{-0.091^\circ}^{+0.094^\circ}$ using data from \textit{Planck} and \textit{WMAP}.  
On the other hand, for anisotropic birefringence that is statistically isotropic, data from several of the current-generation CMB telescopes has been used to extract a measurement of the birefringence power spectrum.  
The results of these studies are summarized in \fref{fig:data}.  
In particular, note that \textit{Planck} data has been analyzed by two different groups using different $\alpha$-estimators, which partly explains the scatter in their results.  
These various measurements in \fref{fig:data} indicate an absence of evidence for anisotropic cosmological birefringence at the level of $\sim 0.1^\circ$ or greater on large angular scales.
Next-generation surveys, such as COrE~\cite{COrE:2011bfs}, LiteBIRD~\cite{Matsumura:2013aja}, Simons Observatory~\cite{SimonsObservatory:2018koc}, CMB Stage IV~\cite{CMB-S4:2016ple}, and PICO~\cite{NASAPICO:2019thw}, expect to deliver measurements of CMB polarization with unprecedented precision.  
These observations will prove to be a powerful probe of cosmological birefringence, improving constraints by $2$ to $3$ additional orders of magnitude~\cite{Pogosian:2019jbt}, and potentially uncovering evidence for birefringence from axion string-wall networks.    
For the time being, the data shown in \fref{fig:data} imposes constraints on the models, which we quantify in the next section.   

\section{Constraints from anisotropic CMB birefringence measurements}
\label{sec:constraints}

In this section we outline the Bayesian inference method have used to search for evidence of birefringence in CMB data. 
For each dataset we approximate the likelihood as a Gaussian and assume vanishing covariance across multipoles (data is available for each multipole up to $\ell = 30$, whereas data above $\ell = 30$ is binned).
We take the log-likelihood to be 
\begin{align}\label{eq:likelihood}
    \ln \mathcal{L}\bigl(C_\ell^\mathrm{obs}\vert \theta\bigr) = \sum_\ell -\frac{1}{2\sigma_\ell^2}\Bigl[C_\ell^\mathrm{obs} - C_\ell^\mathrm{th}(\theta)\Bigr]^2 
    \;,
\end{align}
where $\theta$ represents the model parameters of the theory (listed in \tref{tab:priors} along with the assumed priors), and $\sigma_\ell$ are the uncertainty in the observed values $C_\ell^\mathrm{obs}$. 
We take them to be the error bars in the published birefringence power spectrum plots, which are reproduced in \fref{fig:data}. 
To obtain the posterior for our likelihood and priors, we perform a Markov Chain Monte Carlo (MCMC) simulation using the Metropolis algorithm implemented in the Python package \texttt{PyMC}~\cite{salvatier2016probabilistic}. 
For all data sets, we ran $10$ chains in parallel for at least $5,000$ steps (with some up to even $50,000$ steps depending upon the data set).\footnote{\texttt{PyMC} requires the user to specify the number of tuning steps, used to optimize the sampling algorithm. We used anywhere between $1000-2500$ tuning steps (for different data sets), which were eventually discarded to get all of our final results.} 
We assess their convergence by manually inspecting their trace plots for good mixing, and also ensuring that the Gelman-Rubin statistic $\hat R$~\cite{gelman1992single}, for each model parameter, is close to $1$. 
For each parameter $X \in \theta$ (e.g. $X = \ampl$), our Markov chains satisfy $|\hat R_X - 1| < 0.04$.

\begin{table}[t]
    \centering
    \begin{tabular}{c||c|c|c|c}
    & \text{stable strings} 
    & \text{collapsing string-wall} 
\\ 
    \hline
    \text{amplitude:} 
    & $\ampl \sim \Uniform{-\infty}{\infty}$ 
    & $\ampl \sim \Uniform{-\infty}{\infty}$ 
\\ 
    \text{loop length:} 
    & $\zeta_0 \sim \Uniform{0.1}{1.0}$ 
    & $\zeta_0 \sim \Uniform{0.1}{1.0}$ 
\\ 
    \text{axion mass:} 
    & \text{N/A}
    & $\log_{10}\!\left(\frac{m_a}{\mathrm{eV}}\right) \sim \Uniform{-32.4}{-28.0}$ 
\\ 
    \text{DW number:} 
    & \text{N/A}
    & $\Ndw = 1$
    \end{tabular}
    \caption{\label{tab:priors}The two string-wall network models that we study, their model parameters, and the prior ranges used for MCMC sampling.  The function $\mathrm{U}(a,b)$ denotes a uniform probability density on the interval from $a$ to $b$ and a vanishing probability outside this interval.}
\end{table}

\subsection{Stable string network}
\label{sub:stable_strings_results}

We first study a network of stable hyperlight axion strings.  
The birefringence power spectrum $C_\ell^{\alpha\alpha}$ is calculated using the procedure described in \sref{sub:stable_strings}.  
It is a function of the amplitude parameter $\mathcal{A}^2 \xi_0$ and the loop length parameter $\zeta_0$, which have the priors shown in \tref{tab:priors}.  
We allow the amplitude parameter $\mathcal{A}^2 \xi_0$ to take unphysical, negative values in order to assess the presence of systematic bias in the data; we only use positive values to derive constraints.  

\begin{figure}[t]
	\centering
	\includegraphics{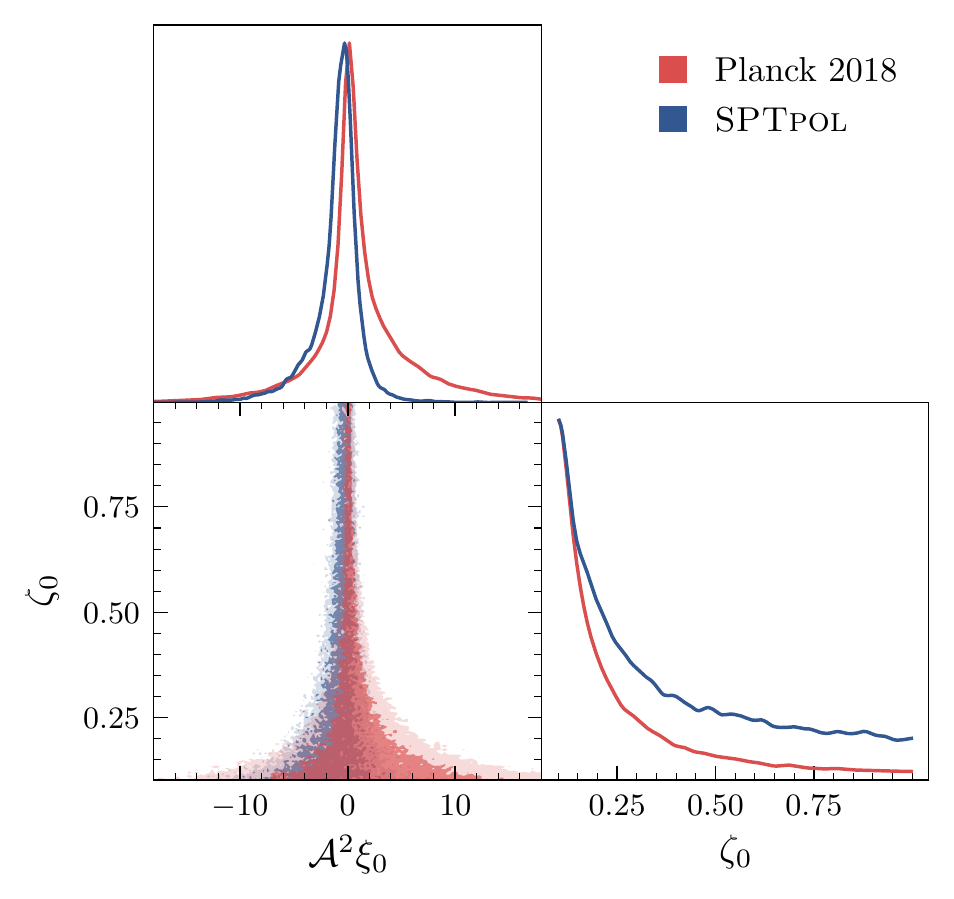} 
	\caption{\label{fig:constraints_stable_string}For the stable string network we show joint posteriors obtained from our MCMC simulations using anisotropic birefringence measurements derived from \textit{Planck} (2018)~\cite{Bortolami:2022whx} and SPT{\sc pol}~\cite{SPT:2020cxx} observations.  Light and dark contours show 95\% and 68\% CL regions respectively. }
\end{figure}

The result of our MCMC sampling is summarized in \fref{fig:constraints_stable_string}, which shows the posterior probability distribution over the amplitude parameter $\ampl$ and the loop length parameter $\zeta_0$. 
This figure illustrates the constraints from \textit{Planck} (2018) ($1 \leq \ell \leq 24$) and SPT{\sc pol} ($75 \leq \ell \leq 525$); constraints from other data sets can be found in \aref{app:different_data}.  
The joint posterior distribution shows a degeneracy direction where $\ampl = 0$, since $C_\ell^{\alpha\alpha}$ becomes independent of $\zeta_0$ when $\ampl = 0$.  
Similarly, the joint posterior broadens toward smaller $\zeta_0$, since $\zeta_0 = 0$ is another degeneracy direction; our prior enforces $0.1 \leq \zeta_0$ (see discussion in \sref{sub:stable_strings}), and the degeneracy at $\zeta_0 = 0$ is not seen on the plot.  
The SPT{\sc pol} data has a very slight preference for $\ampl < 0$ due to a pair of downward fluctuations in the data at $\ell = 120$ and $160$, whereas the \textit{Planck} (2018) data has a wider tail toward $\ampl > 0$ due to a few upward fluctuations at $\ell = 2$, $5$, and $6$.  
Note that the \textit{Planck} (2018) and SPT{\sc pol} measurements have comparable constraining power, even though the SPT{\sc pol} measurements are almost two orders of magnitude more precise.  
The signal $C_\ell^{\alpha\alpha}$ falls off with increasing $\ell$, while the precision of the data improves at a comparable rate up to about $\ell \sim 200$. 
Thus, similar limits are obtained from \textit{Planck} at low $\ell$ and SPT{\sc pol} at high $\ell$ (and also other data sets, with the exception of {\sc Polarbear}; see \aref{app:different_data}).   
Furthermore, since the signal spectrum drops off more quickly than the measurements' precision beyond $\ell \gtrsim 200$, 
we do not expect such high $\ell$ data points to contribute significantly towards our results.

Using the marginalized posterior distribution over the amplitude parameter $\ampl$, we derive 95\% confidence level upper limits for both data sets.  
In doing so, we discard the unphysical parameter space with $\ampl < 0$, and we enforce $\ampl \geq 0$.  
\textit{Planck} (2018) gives $\ampl < 13$ (95\% CL) and SPT{\sc pol} gives $\ampl < 3.7$ (95\% CL).  
The SPT{\sc pol} limit is tighter, partly because the posterior distribution is slightly skewed toward negative amplitudes, and we take only $\ampl > 0$ to derive the limits. 
Since we expect $\mathcal{A} = \mathcal{O}(1)$ from UV model building and $\xi_0 = \mathcal{O}(1-10)$ from string network simulations, these limits are already strongly constraining.

These results are in good agreement with a previous study~\cite{Yin:2021kmx} that calculated the posterior probability distribution over $\ampl$ and $\zeta_0$ and derived constraints on the amplitude $\ampl$ using \textit{Planck} (2015) data~\cite{Contreras:2017sgi}.  
See \aref{app:different_data} for our analysis of the \textit{Planck} (2015) data.  
\Rref{Yin:2021kmx} also presents results for a string network model with a range of loops sizes~\cite{Jain:2021shf}, which we do not repeat here. 
Instead, we provide birefringence constraints on a collapsing string-wall network below.

\subsection{Collapsing string-wall network}
\label{sub:collapsing_string_wall_results}

We now study an axion string-wall network that collapses between recombination and today.  
We calculate the birefringence power spectrum $C_\ell^{\alpha\alpha}$ by following the procedure described in \sref{sub:collapsing_string_wall}.  
Tab.~\ref{tab:priors} shows our priors on the three model parameters: the amplitude parameter $\mathcal{A}^2 \xi_0$, the loop length parameter $\zeta_0$, and the axion mass parameter $m_a$ that controls when the network collapses; we fix $\Ndw = 1$.  
For $m_a \lesssim 10^{-32.4} \ \mathrm{eV}$ the string network has not yet collapsed in the universe today, and we revert back to the analysis of \sref{sub:stable_strings_results}, whereas raising $m_a$ causes the network to collapse earlier.  

\begin{figure}[t]
	\centering
	\includegraphics{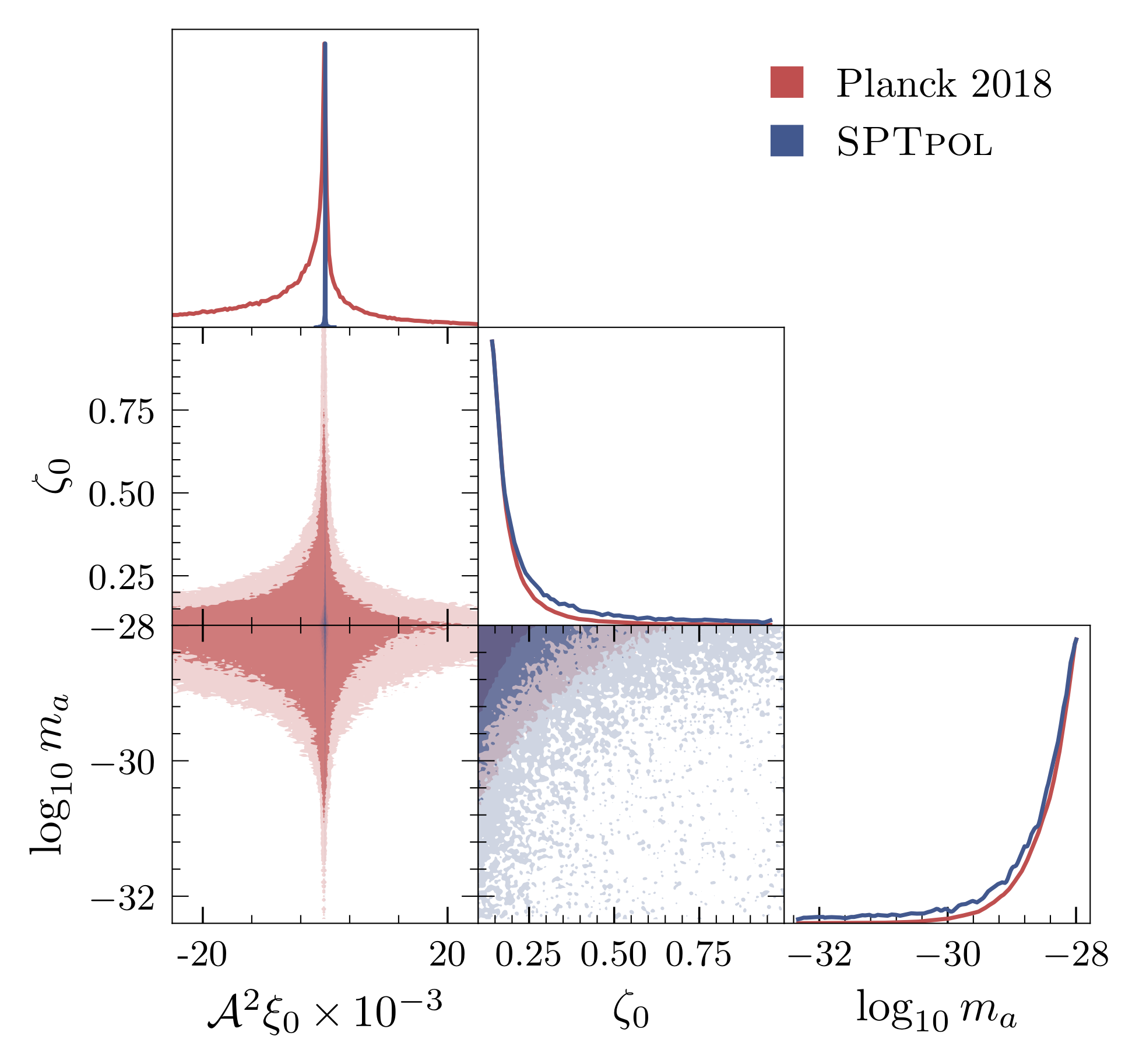} 
	\caption{Joint posteriors for the collapsing string-wall network.  Data and contour shading is the same as \fref{fig:constraints_stable_string}.}
    \label{fig:constraints_collapsing}
\end{figure}

Our results are summarized in \fref{fig:constraints_collapsing}, which shows the marginalized posterior probability distribution over the model parameters.  
The degeneracy direction at $\ampl = 0$ is consistent with the discussion in \sref{sub:stable_strings_results}.  
The data prefers larger values of the axion mass $m_a$, and the marginalized posterior is peaked at the cutoff imposed by the prior $m_a < 10^{-28} \ \mathrm{eV}$.  
This is because the data is consistent with the absence of cosmological birefringence, and raising $m_a$ suppresses power at large angular scales, as seen in \fref{fig:signal_collapse}.  

Using the marginalized posterior distributions we calculate the 95\% CL upper limits on the amplitude parameter, which are found to be $\ampl < 55,\!000$ for \textit{Planck} (2018) and $\ampl < 390$ for SPT{\sc pol}. 
In comparison with our study of the stable string network from \sref{sub:stable_strings_results}, we see that the amplitude limits are weaker here.  
This is partially because raising $m_a$ suppresses power at low $\ell$ and accommodates larger $\ampl$.  
We also note that the \textit{Planck} (2018) limit is weaker than the SPT{\sc pol} limit here by a factor of $\sim 100$.
The strongly scale-dependent power spectrum has $C_\ell^{\alpha\alpha} \propto \ell^0$ at large angular scales, which suppresses the signal in the range of multipoles from $1 \leq \ell \leq 24$ at which the \textit{Planck} (2018) birefringence measurement is available, leading to a weaker limit on the amplitude $\ampl$ as compared with SPT{\sc pol}. 
This observation emphasizes the complementarity between all-sky and ground-based measurements of CMB polarization as probes of cosmological birefringence.  
A detection of anisotropic birefringence on small angular scales ($\ell \sim 100$) without a detection on large angular scales ($\ell \lesssim 10$) would point to a strongly scale-dependent source, and provide evidence for cosmological birefringence from a collapsing axion string-wall network.  

\section{Compatibility with isotropic birefringence measurements}
\label{sec:isotropic}

Aside from searches for anisotropic birefringence, various groups~\cite{Minami:2020odp,Diego-Palazuelos:2022dsq,Eskilt:2022wav,Eskilt:2022cff} have recently analyzed all-sky polarization data to search for evidence of isotropic birefringence.  
In particular, the authors of \rref{Eskilt:2022cff} report a measurement of $\alpha = -{0.342^\circ}_{-0.091^\circ}^{+0.094^\circ}$ (68\% CL) using data from \textit{Planck} and \textit{WMAP}.\footnote{We have adopted the sign convention where a positive birefringence angle induces a counter clockwise rotation in the plane of polarization. Our convention is opposite to the one used in \rref{Eskilt:2022cff}, and we have added a minus sign in reporting their measurement.}  
These analyses provide strong evidence for isotropic birefringence in the CMB at more than $99.9\%$ confidence.  
Here we address the implications of this measurement for birefringence from axion string-wall networks.  

\begin{figure}[p]
	\centering
	\includegraphics{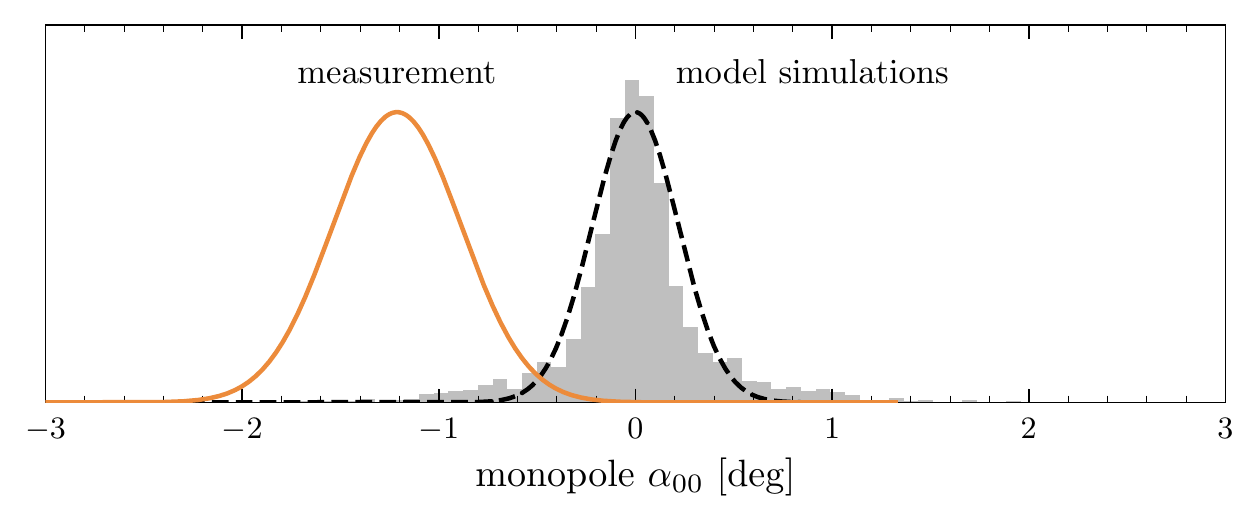}
	\caption{\label{fig:a00_histogram}Distribution over the monopole $\alpha_{00}$ of the birefringence map $\alpha(\nhat)$.  To generate the gray histogram we simulate $1000$ sky maps using the loop crossing model with stable strings for parameters $\mathcal{A} = 1$, $\xi_0 = 0.5$, $\ampl = 0.5$, and $\zeta_0 = 1$. These parameters are chosen to maximize the ``Isotropic BF + SPT{\sc pol}'' distribution shown on \fref{fig:constraints_Komatsu}.  We approximate the simulated distribution by a normal distribution (black-dashed curve) with zero mean and variance $C_0^{\alpha\alpha} = (0.23^\circ)^2$ calculated from the model.  The orange curve shows the measurement of isotropic birefringence from \rref{Eskilt:2022cff}, which we model as a normal distribution with mean $\sqrt{4\pi} \times (-0.342^\circ) = -1.21^\circ$ and standard deviation $\sqrt{4\pi} \times (0.0925^\circ) = 0.328^\circ$.}
\end{figure}

The axion string-wall network produces an anisotropic birefringence signal that is statistically isotropic.  
A general birefringence map can be decomposed onto spherical harmonics as $\alpha(\nhat) = \sum_{\ell,m} \alpha_{\ell m} Y_{\ell m}(\nhat)$, and we are interested in the monopole $\alpha_{00}$, which corresponds to isotropic birefringence.  
Averaging $\alpha_{00}$ over an ensemble of universes gives zero, since positive and negative fluctuations are equally likely.  
Nevertheless, every individual Universe has a nonzero $\alpha_{00}$.  
To illustrate this point, we simulate $1000$ realizations of the birefringence map for the stable string network model, and plot the distribution over the monopole in \fref{fig:a00_histogram}.  
As expected, the mean is close to zero and the variance is approximately $C_0^{\alpha\alpha}$ that we calculate from theory.  
The distribution is clearly non-Gaussian, displaying a tighter central distribution, and moreover the circular features in the sky map imply correlations across modes.  
Nevertheless, a normal distribution with zero mean provides a good approximation.  
We intend to investigate the non-Gaussian behavior further in future work.  

We evaluate the likelihood for isotropic birefringence as follows.  
We treat $\alpha_{00} = S + N$ as the sum of uncorrelated signal and noise terms.  
The signal is modeled as a Gaussian random variable with zero mean and variance $C_0^{\alpha\alpha}$ that we calculate in the loop crossing model. 
The slightly non-Gaussian nature of the $\alpha_{00}$ distribution is neglected for this analysis.  
The noise is modeled as a Gaussian random variable with zero mean and standard deviation $\sigma_0 = \sqrt{4\pi} \times 0.0925^\circ$; this corresponds to the average uncertainty in the isotropic birefringence measurement from \rref{Eskilt:2022cff}, and the factor of $\sqrt{4\pi}$ accounts for the normalization of the spherical harmonics.\footnote{If $\alpha(\nhat) = \sum_{\ell,m} \alpha_{\ell m} Y_{\ell m}(\nhat)$ with $Y_{00} = 1/\sqrt{4\pi}$ then the sky-averaged isotropic birefringence angle is $\bar{\alpha} = \int \! \mathrm{d}^2\nhat \, \alpha(\nhat) / 4\pi = \alpha_{00} / \sqrt{4\pi}$.  We are grateful to Eiichiro Komatsu for pointing out this distinction.}
We extend our log-likelihood~\eqref{eq:likelihood} to include the $\ell=0$ mode in this way, assuming it is uncorrelated with the other multipoles, and we evaluate it at $\alpha_{00} = \sqrt{4\pi} \times (-0.342^\circ)$.  
Note that the sign of $\alpha_{00}$ does not provide any constraint on the model; the monopole $\alpha_{00}$ follows a symmetric distribution with zero mean.  

\begin{figure}[p]
	\centering
	\includegraphics{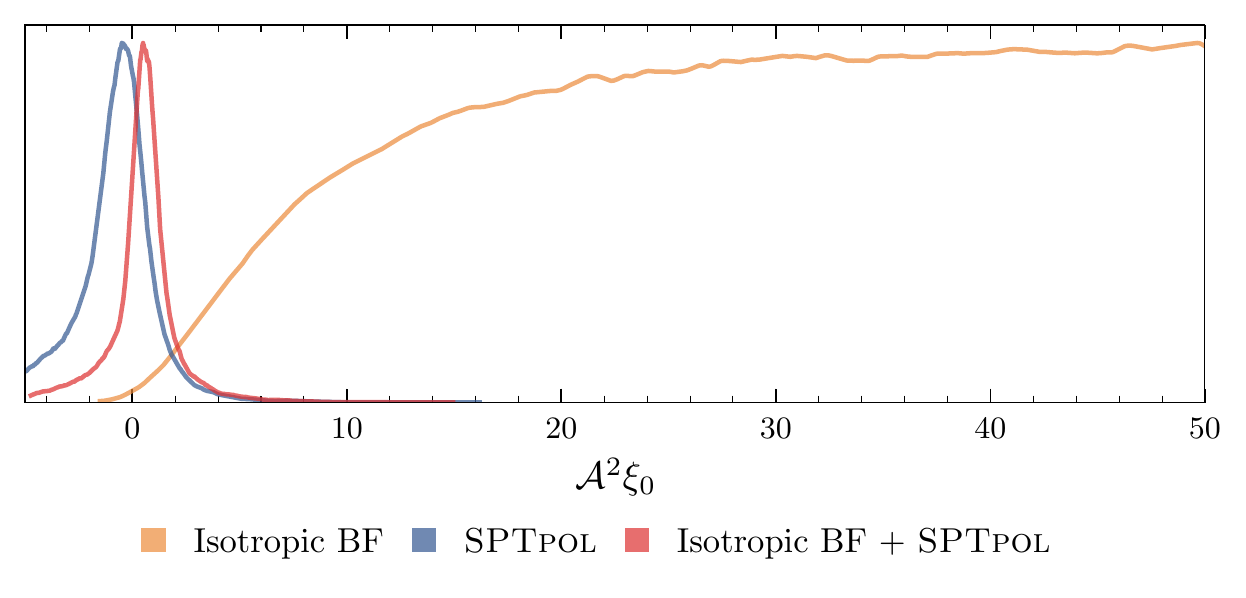}
	\caption{\label{fig:constraints_Komatsu}Assessing the compatibility of isotropic and anisotropic birefringence measurements.  We show the marginalized posterior over the amplitude parameter $\ampl$ for the stable string network model.  The isotropic birefringence measurement favors a nonzero amplitude to fit the monopole $\alpha_{00}$ (orange), whereas the anisotropic measurements using SPT{\sc pol} data constrain the amplitude around zero (blue).  The small overlap of the two distributions illustrates the difficulty in accommodating both measurements from axion-defect-induced birefringence. A joint likelihood combining both measurements (red) favors $\ampl = 0.5 \pm 1.0$ at 68\% CL. }
\end{figure}

We repeat the MCMC analysis and present the results in \fref{fig:constraints_Komatsu}.  
This figure shows the marginalized posterior distribution over the amplitude parameter $\ampl$ for the stable string network model.  
First, taking only the anisotropic birefringence measurements from SPT{\sc pol}, this data is consistent with the absence of an ALP string network, implying $\ampl < 3.7$ (95\% CL), as we also discuss in \sref{sub:stable_strings_results}.  
Second, the isotropic birefringence measurement alone strongly favors the presence of an ALP string network; the posterior is broad, peaking at $\ampl \approx 40-50$ and extending to much larger values (not shown).  
Third, we show the fit to the joint likelihood, which leads to the measurement $\ampl = 0.5 \pm 1.0$ ($68\%$ CL).  

The anisotropic and isotropic birefringence measurements are difficult to reconcile in the context of axion-defect-induced birefringence; this is one of the key results of our work.
The isotropic birefringence measurement favors a large amplitude parameter $\ampl$, which is in conflict with the anisotropic birefringence measurements.  
There is only a small overlap of the posterior distributions in the tail regions.  
Although we present results for SPT{\sc pol} here, the same conclusions can be drawn from the \textit{Planck} (2018) measurement of anisotropic birefringence (or other data sets) instead; however, the wider tail of the \textit{Planck} posterior distribution (see \fref{fig:constraints_stable_string}) leads to a smaller tension.  

We have also performed a similar analysis for the collapsing string-wall network model.  
Since larger $m_a$ suppresses the birefringence signal at low $\ell$, an even larger amplitude $\ampl$ is required to accommodate the isotropic birefringence measurement at $\ell=0$.  
However, this large amplitude comes into sharper tension with the anisotropic birefringence measurements at $\ell > 0$.   

\section{Summary and conclusion}
\label{sec:summary}

In this work we have studied models of axion-like particles that form a network of cosmic strings and domain walls.  
We distinguish four model classes in the parameter space spanned by the axion mass $m_a$ and the domain wall number $\Ndw$: (1) a stable string network that survives in the universe today, (2) a string network that forms domain walls and collapses between recombination and today, (3) a string network that forms stable domain walls between recombination and today, and (4) a string network that forms stable domain walls before recombination. 

We calculate the cosmological birefringence signal that these axion string-wall networks imprint on the polarization pattern of CMB radiation via the usual coupling of the axion-like particles to electromagnetism.  
Using measurements of anisotropic birefringence derived from polarization data taken by various CMB telescopes, we assess the extent to which they are compatible with axion-defect-induced birefringence.  
All of the measurements are consistent with the absence of birefringence from axion string-wall networks, and we derive constraints on the amplitude of the signal.  
Our main results are:  
\begin{itemize}

\item  
For hyperlight ALP masses $m_a \lesssim 3H_0 \simeq 4 \times 10^{-33} \ \mathrm{eV}$, we find that SPT{\sc pol} measurements constrain $\ampl < 3.7$ at 95\% CL (assuming $\ampl \geq 0$) where $\mathcal{A} = - \pi f_a g_{a\gamma\gamma} / \alphaem$ parametrizes the strength of the axion-photon coupling, and $\xi_0$ parametrizes the average total string length in a Hubble volume in units of the Hubble length. In UV extensions of this effective theory, the parameter $\mathcal{A}$ corresponds to an anomaly coefficient, which is model dependent but typically equals an $\mathcal{O}(1)$ rational number. The precise expected value of $\xi_0$ is a matter of some debate, but broadly speaking $\xi_0 = \mathcal{O}(1-10)$.  For instance if $\xi_0 = 30$ then the constraint implies $\mathcal{A} \lesssim 1/4$.  Thus, we conclude that SPT{\sc pol} measurements are already placing meaningful constraints on hyperlight axion-like particles and their UV embedding.

\item  For ALP masses in the range between $3 H_0$ and $3 H_\cmb \simeq 1 \times 10^{-28} \ \mathrm{eV}$ and for $\Ndw = 1$, the anisotropic birefringence signal is predicted to be strongly scale dependent $\ell(\ell+1)C_\ell^{\alpha\alpha} \propto \ell^2$ for small $\ell$ (i.e. on large angular scales the birefringence angles are uncorrelated). This is because the ALP string network develops unstable domain walls when $3H \approx m_a$, causing the string-wall network to collapse and shutting off the source of large-scale birefringence. We find that current measurements of CMB polarization provide no evidence for this signal, which allows us to derive constraints on the axion mass $m_a$ and signal amplitude $\ampl$. Looking forward to future surveys, this distinctive scale-dependent signal provides a compelling target, since its detection would furnish a measurement of the axion mass scale in the range $3 H_0 \lesssim m_a \lesssim 3 H_\cmb$.  It may also be accessible to redshift-dependent probes of birefringence~\cite{Sherwin:2021vgb}.

\item For larger domain wall numbers $\Ndw \geq 2$, the formation of domain walls when $3H \approx m_a$ leads to a stable string-wall defect network.  We argue that the expected birefringence signal is qualitatively equivalent to the case of a stable string network with $m_a \lesssim 3 H_0$.  Thus, we do not perform a separate constraint analysis of these stable string-wall networks, but rather expect out constraints from the stable string network discussed in \sref{sub:stable_strings_results} to carry over.  It is interesting that the stable string-wall network can be consistent with overclosure constraints (depending on $f_a$ and $\Ndw$) even for ALP masses as large as $m_a \sim 10^{-20} \ \mathrm{eV}$, which is the usual range of ultralight bosonic dark matter. This observation provides motivation to study the connections between ultralight ALP dark matter, astrophysical constraints, and cosmological signatures.  

\item For hyperlight ALP masses, we find that the measurements of anisotropic birefringence derived from ground-based telescopes such as ACT{\sc pol} and SPT{\sc pol}, currently provide the strongest constraints on axion string-wall networks. A plausible explanation is as follows.  The birefringence power spectrum $\ell(\ell+1)C^{\alpha\alpha}_\ell$ peaks around $\ell_p \sim 40/\zeta_0$ before decreasing again at larger multipoles. This translates to $C^{\alpha\alpha}_{\ell_p} \sim (\Acal^2\xi_0\,\zeta_0^2)\,(4\times 10^{-5})$. Since the current ACT{\sc pol} and SPT{\sc pol} measurements have data points around this region with strongest precision ($\sigma_\ell \lesssim 10^{-4} \ \mathrm{deg}^2$), they end up providing the most stringent constraints. 

\item  
Among the various data sets that we consider, we find that anisotropic birefringence measurements derived from SPT{\sc pol} data yield the strongest constraints on axion-defect-induced birefringence.  
This is partly because of two downward fluctuating data points at $\ell = 120$ and $\ell = 160$, which skew the amplitude distribution towards negative values, leading to a tighter 95\% CL upper limit on $\ampl$ assuming $\ampl \geq 0$.  

\item  We assess the extent to which recent measurements of a nonzero isotropic birefringence are consistent with constraints on anisotropic birefringence in the context of an axion-defect-induced signal.  We find that it is somewhat difficult for a stable string network (or stable string-wall network) to induce a birefringence signal that is compatible with the isotropic measurement and the lack of an anisotropic signal. 
The situation is further exacerbated for the collapsing string-wall network models due to additional reduced power on large angular scales.

\end{itemize}

Finally, let us remark on potential directions for extending the computational framework in which our results have been derived.  
The rich dynamics of a topological defect network present a challenge toward deriving phenomenological observables.  
In the work presented here, we have used the loop-crossing model (see \sref{sec:birefringence}) to reduce the complex network down to a manageable number of degrees of freedom with which we can calculate a birefringence signal.  
The loop crossing model does not capture certain features that an axion-string wall network is expected to exhibit: the finite duration of domain wall formation and network collapse around $3H \sim m_a$ (for $\Ndw=1$), or transition into a new scaling solution (for  $\Ndw\geq2$); and the gradual change in the axion field nearby to a string loop. 
This last feature is expected to impact the low-$\ell$ power spectrum, since the axion field does not change by the full asymptotic amount of $\sim 2\pi f_a$, leading to only $\alpha < \mathcal{A} \alphaem$ during the last few e-foldings. 
Or in the case of stable walls, there aren't $\sim \Ndw$ wall crossings in these last few e-foldings before the CMB light reaches us in the present. 
It is important to understand how each of these features affects the birefringence signal in order to derive robust limits on axion string-wall networks from next-generation surveys.  

\begin{acknowledgments}
We are grateful to Anthony Challinor, Richard Battye, Steve Gratton, Wayne Hu, Toshiya Namikawa, Ippei Obata, and Blake Sherwin for valuable discussions.   We would particularly like to thank Joel Meyers and Levon Pogosian for guidance in the early stages of this project, and Eiichiro Komatsu for his timely feedback as we were finalizing the project. 
R.H, M.J., and A.J.L.~are supported in part by the National Science Foundation under Award No.~PHY-2114024. MA is supported by a DOE grant DOE-0000250746. 
\end{acknowledgments}

\appendix

\section{Simulating loop crossing model}
\label{app:simulation}

Here we present a step-by-step procedure of how we simulate the loop crossing model to generate birefringence maps using HEALpix~\cite{Gorski:2004by}.

\begin{enumerate}[label*=\arabic*.]

    \item {\bf Initialize HEALPix map.} We begin by making a HEALpix map with pixel parameter $N_{\mathrm{side}} = 2048$, and initialize a null array of length  $N_{\mathrm{pix}} = 12 N_\mathrm{side}^2$.
    
    \item {\bf Choose time steps.} For time evolution, we pick redshifts in logarithmic intervals from recombination $z_\cmb = 1100$ till the present $z = 0$, in the following manner
    \begin{align}
        1 + z_n = (1 + z_\cmb) \left(\frac{1 + z_\mathrm{final}}{1 + z_\cmb} \right)^{(n-X) / N_\mathrm{steps}}\,.
    \end{align}
    Here, the index $n$ goes from $1$ to $N_\mathrm{steps} = 28$, and $X$ is a random variable sampled uniformly between $-1/2$ and $1/2$ at the start of the simulation (subsequent steps then use the same value). This ensures each simulation samples different redshift steps over many simulations. This reproduces a continuous evolution, when averaged over a large ensemble of simulations. 
    
    \item {\bf Run simulation.} For each redshift step $z_n$:
    \begin{enumerate}[label*=\arabic*.]
        \item {\bf } We populate the CMB light cone (from $z_{n-1}$ to $z_{n}$) with circular loops, all having the same dimensionless radius $\zeta$.
        The number of loops is determined by calculating the average number of loops $N$ and randomly selecting each pixel to be the center of a loop with probability $N / N_{\mathrm{pix}}$. Assuming the network is in scaling, the average number density is 
        \begin{align}
            n = \frac{\rho}{E} = \frac{\xi_0 H^3}{2\pi \zeta}\,.
        \end{align}
        Here $\rho = \xi_0 \mu H^2$ is the energy density of the scaling network, $E = \mu 2\pi\zeta_0 H^{-1}$ is the energy of a circular string loop with radius $\zeta_0 H^{-1}$ and tension $\mu$. The average number of loops on the CMB light cone from $z_{n-1}$ to $z_n$ is therefore
        \bes{
            N &= \int \! \dd V_\mathrm{light cone}\, (\mathrm{number\ density})\\
            &=  \int_{z_n}^{z_{n-1}}\!\dd z \, \frac{\xi_0 (a H)^3}{2\pi \zeta_0}\ 4\pi s(z)^2 H^{-1}(z)\,.
        }
        Here $s(z)$ is the comoving distance from an observer on Earth to the centers of loops through which CMB photons crossed at redshift $z$.
        
        \item Out of $N_{\mathrm{pix}}$, we randomly select $N_T$ pixels on the sphere for loop centers. Every loop is given a random orientation and assigned a random winding number $w$ (equal to $+1$ or $-1$ with equal probability). For uniformly oriented loops, $\cos{\Theta} \sim \mathrm{U}(0,1)$ and $\Phi \sim \mathrm{U}(0,2\pi)$, where $(\Theta,\Phi)$ are polar and azimuthal angles measured relative to the normal vector of the loop's center pixel. 
        
        \item After populating the CMB light cone with $N_T$ loops, for every loop we find the region/pixels bounded by its spherical projection onto the sphere. (The function \texttt{healpy.query\_polygon()} was used for this purpose).
        All the pixels in the region are assigned the value $w\Acal\alphaem$ where $w$ is the winding number of the loop.
         
        \item We repeat this procedure until $n = N_\mathrm{steps}$, and perform many such simulations for ensemble averaging.
    \end{enumerate}
\end{enumerate}

\section{Statistical estimator for anisotropic birefringence}
\label{app:alpha_estimator}

This appendix includes supplementary material regarding: the effect of birefringence on the CMB, a statistical estimator that may be used to measure birefringence from CMB polarization data, and a demonstration of how the estimator works using simulated data.  

The effect of birefringence on the CMB is most easily understood in position space where $\nhat$ is a unit vector at some point on the sky.  
If $\tilde{T}(\nhat)$, $\tilde{Q}(\nhat)$, and $\tilde{U}(\nhat)$ denote the would-be CMB temperature and polarization sky maps in the absence of birefringence, then an anisotropic birefringence angle $\alpha(\nhat)$ mixes $Q$- and $U$-mode polarization~\cite{Zaldarriaga:1996xe}, giving rise to the observable sky maps
\begin{subequations}\label{eq:TQU-transform}
\begin{align}
     T &= \tilde T \\
     Q &= \tilde Q \cos 2\alpha - \tilde U \sin 2\alpha \\ 
     U &= \tilde Q \sin 2\alpha + \tilde U \cos 2\alpha 
     \;.
\end{align}
\end{subequations}
If $\tilde{T}$, $\tilde{Q}$, and $\tilde{U}$ were known, then \eref{eq:TQU-transform} would allow $\alpha$ to be extracted from measurements of $T$, $Q$, and $U$.  
Of course, the would-be temperature and polarization anisotropies of the CMB cannot be calculated (nor measured), but the \emph{statistical properties} of these fields are calculable.  
This observation motivates one to define a birefringence estimator, which reproduces the true birefringence as a statistical average.  

Several statistical estimators of anisotropic birefringence have been proposed in the literature. Here we discuss a particular set of estimators that have been used in recent CMB data analyses. These are similar to the quadratic estimators proposed by Hu and Okamoto~\cite{Hu:2001kj} for studies of CMB weak lensing, building on which the authors of \rref{Kamionkowski:2008fp,Yadav:2009eb,Gluscevic:2009mm} constructed another set of estimators for studies of CMB birefringence.

Following the notation of \rref{Yin:2021kmx}, the birefringence estimators in the flat-sky approximation~\cite{Yadav:2009eb} can be written as 
\begin{align}
    \hat{\alpha}_{XY}(\mathbf{L}) &= \lambda_{XY}(\mathbf{L})\int \! \! \frac{\dd^2 l_1}{(2\pi)^2} X(\lvec_1) Y^\ast(\lvec_2) F_{XY}(\lvec_1, \lvec_2) \big\vert_{\lvec_2=\lvec_1-\mathbf{L}} \label{eq:HO_estimator}
\end{align}
where $X$ and $Y$ stand for temperature $T$, parity-even $E$-mode polarization, or parity-odd $B$-mode polarization.  
In fact there are five different birefringence estimators corresponding to the choice of $XY \in \{ EE, BB, TE, TB, EB \}$ (since $XY=TT$ is trivial).  
In the flat-sky approximation, $\Lvec, \lvec_1, \lvec_2 \in \mathbb{R}^2$ are the analogs of the spherical harmonic integer indices $(\ell,m)$, and \eref{eq:HO_estimator} assumes $\Lvec \neq 0$.  
The estimator integrates over $X(\lvec)$ and $Y(\lvec)$, which represent the observed temperature or polarization maps, weighting them by the mode coupling coefficients 
\begin{align}
    F_{XY}(\lvec_1, \lvec_2) & = 
    \begin{cases}
    \frac{f_{XY}(\lvec_1, \lvec_2)}{(1+\delta_{XY})C_{l_1}^{XX}C_{l_2}^{YY}} & , \quad XY \neq TE \\ 
    \frac{C_{l_1}^{YY} C_{l_2}^{XX} f_{XY}(\lvec_1, \lvec_2)
    - C_{l_1}^{XY}C_{l_2}^{XY} f_{XY}(\lvec_2, \lvec_1)}{C_{l_1}^{XX}C_{l_2}^{YY}C_{l_2}^{XX}C_{l_1}^{YY} - (C_{l_1}^{XY} C_{l_2}^{XY})^2} & , \quad XY = TE
    \end{cases} 
    \;.
\end{align}
Here $C_\ell^{XY}$ are the predicted power spectra (or cross-correlation spectra if $X \neq Y$), $\delta_{XY}$ is the Kronecker delta, and the response functions $f_{XY}(\lvec_1, \lvec_2)$ are given in \tref{tab:response_functions}.  
Finally the normalization coefficient 
\begin{align}
    \left[\lambda_{XY}(\mathbf{L})\right]^{-1} & = \int \! \! \frac{\dd^2 l_1}{(2\pi)^2} f_{XY}(\lvec_1, \lvec_2) F_{XY}(\lvec_1, \lvec_2) \big\vert_{\lvec_2=\mathbf{L} - \lvec_1} 
\end{align}
ensures that $\hat{\alpha}_{XY}(\Lvec)$ is an unbiased estimator.  

\begin{table}[t]
    \begin{center}
        \begin{tabular}{|c|c|}
            \hline
            $XY$ & $f_{XY}(\lvec_1, \lvec_2)$ \\
            \hline
            $TT$ & $0$ \\
            $TE$ & $-2\tilde{C}_{l_1}^{TE}\sin 2\varphi_{12}$ \\
            $TB$ & $2\tilde{C}_{l_1}^{TE}\cos 2\varphi_{12}$ \\
            $EE$ & $-2\bigl(\tilde{C}_{l_1}^{EE}-\tilde{C}_{l_2}^{EE}\bigr)\sin 2\varphi_{12}$ \\
            $EB$ & $2\bigl(\tilde{C}_{l_1}^{EE}-\tilde{C}_{l_2}^{BB}\bigr)\cos 2\varphi_{12}$ \\
            $BB$ & $-2\bigl(\tilde{C}_{l_1}^{BB}-\tilde{C}_{l_2}^{BB}\bigr)\sin 2\varphi_{12}$ \\
            \hline
        \end{tabular}
    \end{center}
    \caption{\label{tab:response_functions}Response functions. Here $\varphi_{ij} = \varphi_{\lvec_i} - \varphi_{\lvec_j}$ where $\cos \varphi_{\lvec} \equiv \nhat\cdot\hat\lvec$. In \texttt{symlens} $\cos{2\varphi_{12}}$ is represented symbolically by \texttt{symlens.cos2t12}.}
\end{table}

We seek to demonstrate how the estimator~\eqref{eq:HO_estimator} reconstructs a known birefringence map $\alpha(\nhat)$ from an ensemble of simulated temperature and polarization maps.  
This is done with the following procedure. 

\begin{enumerate}
    
    \item We begin by calculating the angular power spectra $\tilde C_\ell^{TT}$, $\tilde C_\ell^{EE}$, and $\tilde C_\ell^{TE}$ using CAMB~\cite{Lewis:1999bs} (we take $\tilde C_\ell^{BB} = \tilde C_\ell^{TB} = \tilde C_\ell^{EB} = 0$). Assuming Gaussian fluctuations, we construct all-sky temperature and polarization maps, $\tilde T(\nhat)$, $\tilde Q(\nhat)$, and $\tilde U(\nhat)$ using \texttt{healpy} (the Python wrapper for \texttt{HEALPix}).
    
    \item Next, we generate a birefringence map $\alpha(\nhat)$ using the procedure outlined in \aref{app:simulation}; since this is just a proof of principle demonstration, the parameters of the string network model are not particularly important.  
    
    \item Using the simulated birefringence map, we transform the temperature and polarization maps according to \eref{eq:TQU-transform}, and convert $Q$ and $U$ maps to $E$ and $B$ maps.
    
    \item In order to apply the flat-sky estimator, we isolate some small section of the all-sky maps and port it into the \texttt{pixell}~\cite{pixell} software. 
    
    \item With the chosen small section, we calculate the birefringence estimator $\hat{\alpha}_{EB}(\Lvec)$. We use \texttt{symlens}~\cite{symlens} to evaluate the integrals in \eref{eq:HO_estimator}.
    
    \item Finally we perform an inverse 2D Fourier transform on $\hat \alpha_{EB}(\Lvec)$ to obtain the reconstructed birefringence map $\hat{\alpha}_{EB}(\nhat)$.

\end{enumerate}

\begin{figure}[t]
	\centering
	\includegraphics[width=0.3\textwidth]{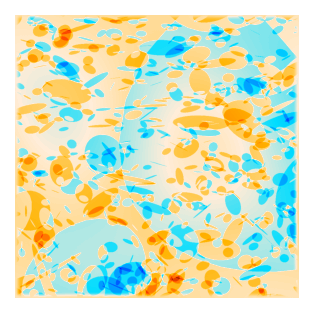}\hspace{-1px}
	\includegraphics[width=0.3\textwidth]{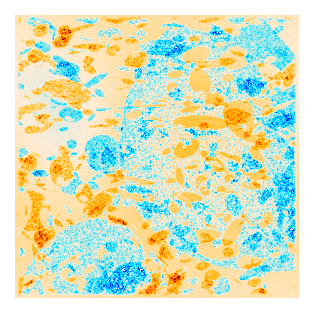}\hspace{-1 px}
	\includegraphics[width=0.3\textwidth]{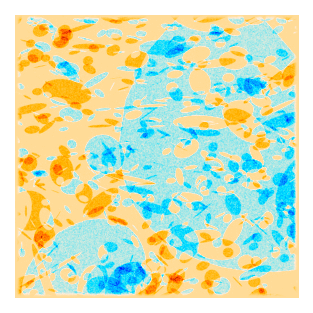}
	\caption{\label{fig:alpha_est_pipeline}A demonstration of how the statistical estimator $\hat \alpha_{EB}(\nhat)$ from \eref{eq:HO_estimator} reconstructs a birefringence map $\alpha(\nhat)$.  \emph{Left:} The `true' birefringence map $\alpha(\nhat)$.  \emph{Middle:} The reconstructed birefringence map $\hat \alpha_{EB}(\nhat)$ obtained from a single realization of the CMB temperature and polarization maps. \emph{Right:} Reconstructed birefringence map $\hat \alpha_{EB}(\nhat)$, averaged over a suite of $10$ realizations. Our implementation of the estimator in this figure introduces a multiplicative bias (not perceptible here) that scales inversely with the map width.
	}
\end{figure}

\Fref{fig:alpha_est_pipeline} shows the result of the above procedure. The left panel shows the `true' birefringence map $\alpha(\nhat)$, which acts on the CMB polarization anisotropies according to \eref{eq:TQU-transform}. The middle and right panels show the reconstructed birefringence map $\hat{\alpha}_{EB}(\nhat)$ using only $1$ and an average over $20$ realizations of the CMB, respectively. Upon averaging over many CMB realizations, the estimator should converge to the true birefringence map (for an unbiased estimator).  We note that even a single CMB realization leads to a reliable reconstruction that captures many qualitative properties of the true map, e.g. scale and shape of loop-like features.  

\section{Alternative birefringence data}
\label{app:different_data}

Measurements of anisotropic cosmological birefringence are available from the various CMB telescopes; see \fref{fig:data}.  
Using each of these data sets individually, we derive constraints on the axion string-wall network models.  
Our results are summarized in this appendix.  

\begin{figure}[t]
	\centering
	\includegraphics{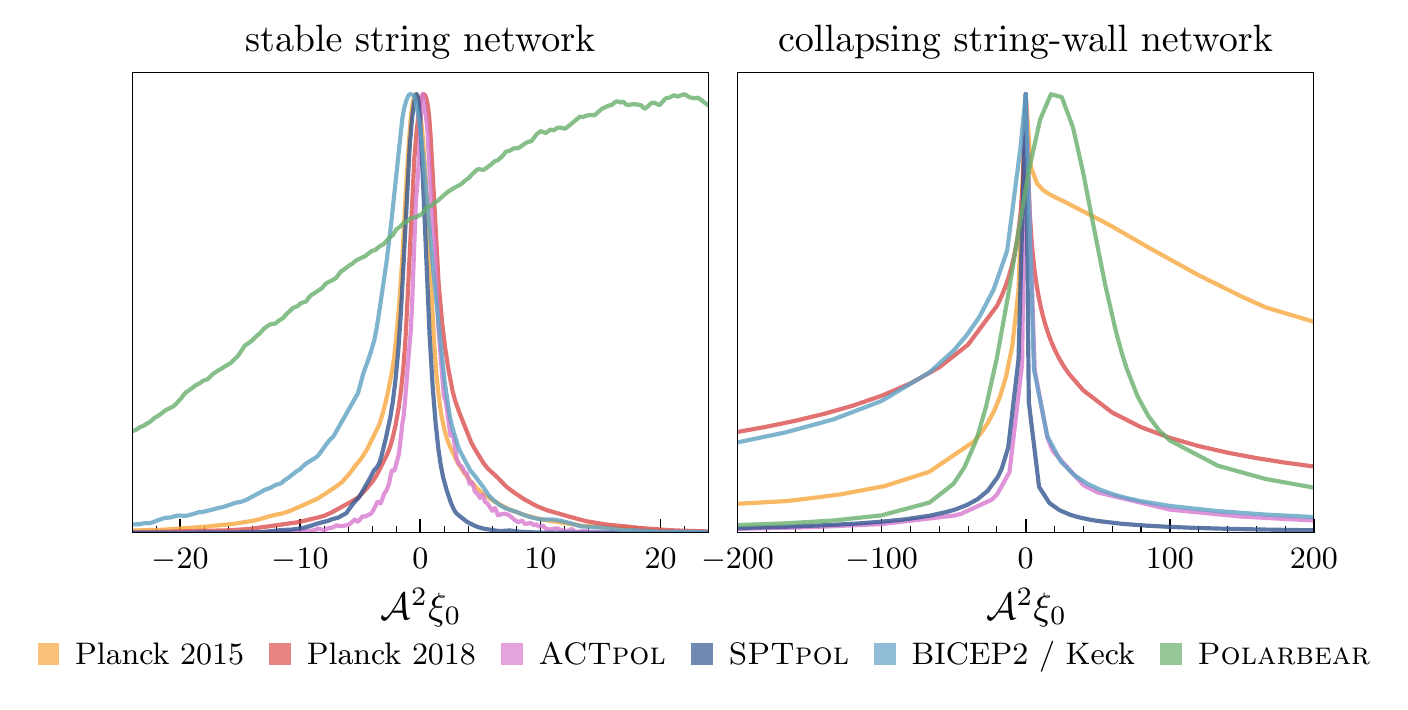}
	\caption{\label{fig:constraints_all_data}Marginalized posterior on the amplitude parameter $\ampl$ for a network of stables strings (left panel) and a collapsing string-wall network (right panel). 
	}
\end{figure}

\Fref{fig:constraints_all_data} shows the marginalized probability distribution over the amplitude parameter $\ampl$ for the stable string network (left panel) and the collapsing string-wall network (right panel).  
Each curve corresponds to a measurement of anisotropic birefringence using data from a different CMB telescope; see~\fref{fig:data}.  
The corresponding 95\% CL upper limits on the amplitude parameter (assuming $0<\ampl$) are summarized in~\tref{tab:ampl_limits}.  

\begin{table}[t!]
    \centering
    \begin{tabular}{c||c|c}
    & \text{stable strings} & \text{collapsing string-wall} \\ 
    \hline
    \text{\textit{Planck} (2015):} & $\ampl < 13$ & $\ampl < 55,000$ \\ 
    \text{\textit{Planck} (2018):} & $\ampl < 13$ & $\ampl < 18,000$ \\ 
    \text{ACT{\sc pol}:} & $\ampl < 7.1$ & $\ampl < 1,100$ \\
    \text{SPT{\sc pol}:} & $\ampl < 3.7$ & $\ampl < 390$ \\  
    \text{BICEP2/\textit{Keck}:} & $\ampl < 11$ & $\ampl < 3,200$ \\ 
    \text{{\sc Polarbear}:} & $\ampl < 81$ & $\ampl < 3,300$ 
    \end{tabular}
    \caption{\label{tab:ampl_limits}Upper limits at 95\% CL on the amplitude parameter $\ampl$ derived from measurements of anisotropic birefringence using data from various CMB telescopes.  }
\end{table}

From these results, one can see that each of the data sets is consistent with $\ampl=0$ at the $1\sigma$ level, corresponding to the absence of axion-defect-induced birefringence.  
SPT{\sc pol} provides the strongest constraints on the amplitude of the signal, for both the stable string network model and the collapsing string-wall network model.  
An upward fluctuation in the {\sc Polarbear} data~\cite{POLARBEAR:2015ktq} leads to a $\sim 1\sigma$ preference for $\ampl > 0$, whereas several downward fluctuations in the BICEP2/\textit{Keck Array} data~\cite{BICEP2:2017lpa} broaden the distribution toward negative amplitudes.  
For the collapsing string-wall network, the \textit{Planck} data is less constraining, consistent with the discussion in \sref{sec:constraints}.  

Using a birefringence measurement~\cite{Contreras:2017sgi} derived from \textit{Planck} (2015) data, we find that the network of stable strings is constrained by $\ampl < 13$ at 95\% CL.  
This particular data set was also analyzed by another recent study~\cite{Yin:2021kmx}, and our results are in good agreement.  
Note that \rref{Yin:2021kmx} presents a constraint $\ampl < 8.0$ at 95\% CL, which is derived allowing $\ampl < 0$.  
To compare with our result, we digitize the marginalized posterior from fig.~5 of \rref{Yin:2021kmx} and calculate $\ampl < 12.6$ at 95\% CL when imposing $\ampl > 0$, which is in excellent agreement with our result.

\bibliographystyle{JHEP}
\bibliography{references}

\end{document}